\documentclass[floatfix,reprint,amsmath,amssymb,aps,]{revtex4-2}
\usepackage{xcolor}
\usepackage{graphicx}
\usepackage{dcolumn}
\usepackage{bm}
\usepackage{hyperref}
\usepackage{bbold}
\usepackage{yhmath}
\usepackage[english]{babel}
\usepackage[utf8]{inputenc}

\begin{document}

\preprint{APS/123-QED}

\title{Generation of phonon quantum states and quantum correlations among single photon emitters in hexagonal boron nitride}

\author{Hugo Molinares$^1$, Fernanda Pinilla$^2$,  Enrique Mu\~noz$^{3,4}$, Francisco Mu\~noz$^{2,5}$, Vitalie Eremeev$^{6,7, }$}  
\email{vitalie.eremeev@udp.cl}

\affiliation{$^1$ Departamento de Ciencias Físicas, Universidad de La Frontera, Casilla 54-D, Temuco 4780000, Chile.}

\affiliation{$^2$Departamento de F\'isica, Facultad de Ciencias, Universidad de Chile, Santiago, Chile.}

\affiliation{$^3$Institute of Physics, Pontificia Universidad Cat\'olica de Chile, Santiago, Chile.}

\affiliation{$^4$Center for Nanotechnology and Advanced Materials CIEN-UC,
Avenida Vicu\~a Mackenna 4860, Santiago, Chile}

\affiliation{$^5$Center for the Development of Nanoscience and Nanotechnology (CEDENNA), Santiago, Chile.}

\affiliation{$^6$ Instituto de Ciencias B\'asicas, Facultad de Ingenier\'ia y Ciencias, Universidad Diego Portales, Av.
 Ejercito 441, Santiago, Chile }
 
\affiliation{$^7$Institute of Applied Physics, Academiei 5, MD-2028, Chi\c{s}in\u{a}u, Moldova.}

\date{\today}

\begin{abstract}
Hexagonal boron nitride exhibits two types of defects with great potential for quantum information technologies: single-photon emitters (SPEs) and one-dimensional grain boundaries hosting topologically-protected phonons, termed as {\it{topologically-protected phonon lines}} (TPL). Here, by means of a simple effective model and density functional theory calculations, we show that it is possible to use these phonons for the transmission of information. Particularly, a single SPE can be used to induce single-, two- and qubit-phonon states in the one dimensional channel, and \textit{(ii)} two distant SPEs can be coupled by the TPL that acts as a waveguide, thus exhibiting strong quantum correlations. We highlight the possibilities offered by this material-built-in nano-architecture as a phononic device for quantum information technologies.\end{abstract}

\maketitle

\section{\label{sec:level1}Introduction}

Phonon-based quantum devices have attracted recent attention since, among other advantages over photons, phonons offer richer alternatives for coupling with several solid-state quantum systems\cite{barfuss2015strong,Bhattacharya_2021,whiteley2019,Errando_2021}. There are proposals of systems comprising single photon emitters (SPEs), such as color centers in diamond, coupled by a one dimensional waveguide\cite{Peyskens_2019,PhysRevA.101.022311,PhysRevLett.117.015502}. These reports show a promising way to entangle the SPEs. Also more exotic phenomena, such as phonon-phonon correlations\cite{Shen21} or phonon blockade\cite{PRA.93.063861,photonics9020070,PRApp.17.054004} have been explored.

Recently, a two-dimensional system, hexagonal boron nitride (hBN), has attracted the attention as a material for quantum technologies\cite{Abdi2017, Abdi2019} due to the presence of SPEs covering the visible spectrum\cite{Tran2016,Vasconcellos_2022,grosso2017}. Despite the vast differences among these emitters, most of them share a common feature in their photoluminiscence spectrum: well-separated and sharp phonon replicas, at $\sim 160-180$ meV\cite{jero216,Wigger_2019}.

Additionally, hBN has been predicted to host one-dimensional topologically-protected phonons\cite{Jiang2018,LiXi2021} that do not experience dissipation and hence may preserve quantum correlations over comparatively long distances. Moreover, such phonon modes are immune to most types of disorder, as long as the relevant symmetry is protected (see Sec.~\ref{sec:TPL}). They seem ideal for playing the role of a phononic waveguide, able to keep coherence at long distances.

In this article, we propose a system consisting of a layer of hBN hosting two SPEs coupled by a topologically protected phonon channel, see Fig.~\ref{fig1}a. Our theoretical analysis is based on the combination of the spin-boson model to capture the electron-phonon coupling and an external laser pumping, density functional theory to obtain the suitable microscopic parameters, and the solution of the Master Equation (including dissipation with a thermalized phonon bath). Our analysis is focused on the following configurations: (\textit{i}) single-, two- and three-phonon Fock states along the phonon line, (\textit{ii}) phonon-qubit state, and (\textit{iii}) quantum correlations between the SPEs. We will start in Sec.~\ref{sec:model} by introducing the underlying physical system, its components and the effective Hamiltonian describing the interaction. Then, in Sec.~\ref{sec:results}, we will show quantized single- and two-phonon states along the channel, induced by the SPEs and phonon-mediated correlations among the SPEs. We shall discuss our results in Sec.~\ref{sec:disc}, by comparing with an experimental simulation on the Quantum IBM. Sec.~\ref{sec:methods} summarizes our \textit{ab initio} methods. 
Finally, we will mention our conclusions in Sec.~\ref{sec:concl}.

\section{Generalized model}
\label{sec:model}

\subsection{Topologically-protected phonon line in hBN}
\label{sec:TPL}
\begin{figure}
    \centering
    \includegraphics[width=\columnwidth]{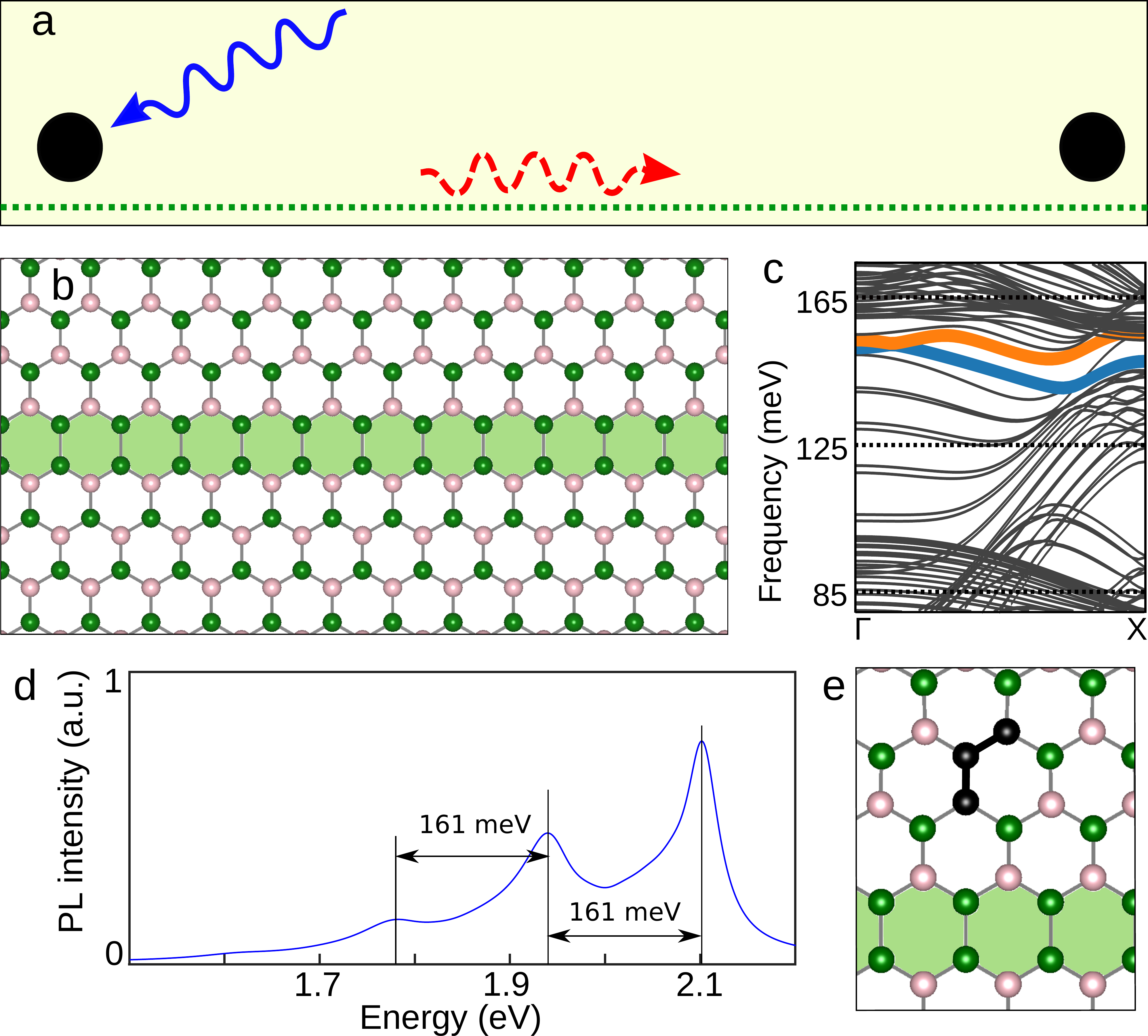}
    \caption{(a) Scheme of a grain boundary (green dashed line) coupling two SPEs (black circles) by means of a topologically-protected phonon mode (red curly arrow). Here, the SPEs can be excited by a laser (blue arrow). (b) Atomic representation of the grain boundary hosting the topologically protected phonons in h-BN, where each color represents a different atomic species. (c) Phonon subbands along the grain boundary, the topologically protected modes are highlighted by blue and orange. (d) Simulation of the photoluminiscence spectra of a typical SPE in hBN (C$_2$C$_N$). (e) Model of a C$_2$C$_N$ defect close to the grain boundary. }
    \label{fig1}
\end{figure}

The unit cell of h-BN consists of two triangular sublattices formed by B and N atoms, respectively. At grain boundaries where the $Z_2$ symmetry is locally broken by the presence of a zig-zag edge of identical atomic species (see Fig.~\ref{fig1}b), a topologically-protected phonon line (TPL) arises\cite{LiXi2021,Jiang2018,Jiang2018b}. Breaking this lattice symmetry globally produces a well known gap-opening effect in the electronic spectrum of single-layer graphene\cite{Kumar_2019,Das_2016}. Here, the mechanism behind the emergence of the TPL seems analogous to the (electronic) valley Hall effect in bilayer graphene\cite{Ju2015,zhang2011,Yao2009,Jung2011,Munoz2016}, which has been applied to other excitations in hexagonal lattices\cite{Munoz2020,Pal_2017}. This effect requires: \textit{(i)} two \textit{physically different} sublattices, in the sense that this difference has to open a band gap in the excitation spectrum, and \textit{(ii)} the sublattices must be interchanged in some region. In the case of the valley Hall effect in bilayer graphene, a gate potential gives a different potential to both sublattices, opening a band gap, and a local inversion of the gate potential can interchange the sublattices, resulting in topological states at the regions where this inversion of voltage occurs. Similarly, in h-BN, the different masses of B and N atoms induces a band gap in the phonon spectrum (\textit{i.e.} lifting a degeneracy at the K point). The remaining ingredient is switching both sublattices, which happens at certain grain boundaries (Fig.~\ref{fig1}b). There are two types of these boundaries, B--B and N--N, in the figure only the B--B is shown, but the localized phonons of each boundary are similar. The phonon subbands of a cell with a pair of -well-separated- grain boundaries, B--B and N--N, are shown in Fig.~\ref{fig1}c. The TPL have an energy centered at $\sim 160$ meV.

\subsection{Single photon emitters in hBN}
\label{sec:spes}
Most SPEs in h-BN are associated with substitutional carbon defects~\cite{Auburber2021,mendelson2021,Jara2021,pinilla2022carbonbased}. Depending on the atomic detail of the actual defects, the emission (zero-phonon line) can range from the near infrared to near ultraviolet energies~\cite{Wigger_2019,Jara2021,Tran2016}. Unlike color centers in diamond, the phononic sideband of the photoluminescence (PL) spectrum shows very prominent replicas with a shift of $\approx 160-180$ meV respect to the zero-phonon line~\cite{Wigger_2019,jero216,Jara2021}. They correspond to bond stretching modes, due to the different geometry of the ground and excited states of the SPEs, respectively. Fig.~\ref{fig1}d shows a typical PL spectrum~\cite{Jara2021,Li2022,jero216}, with a zero-phonon line at $\approx 2.1$ eV. The energy of the phononic replicas is practically independent on the details of SPE~\cite{Wigger_2019}. There exist a very different kind of SPEs in hBN which are based on vacancies, they have much softer and less defined phonon replicas in their PL spectrum~\cite{ivady2020ab,Kianina2020}. Along this article we will ignore these vacancy-related SPEs.

\subsection{Coupling between the TPL and a SPE}
\label{sec:coupling}

The energy of the phonons resonating with the optical transition of a SPE is similar to that of the TPL. If the difference between both energies is small, the SPE's phonons could interact with the TPL, see Fig.~\ref{fig1}e. If a second SPE is close to the TPL, both SPEs could be correlated by the TPL -which then plays the role of a waveguide- as depicted in Fig.~\ref{fig1}a. Being dissipationless, the TPL can -in principle- link SPEs regardless of the distance among them, unlike typical phonons which are subject to umklapp scattering processes. Also, thanks to its inherent topological protection, this channel should be resistant to most types of scattering (\textit{e.g.} lattice phonons), making it ideal for the transfer of information, even at high temperature. However, the TPL is not robust to disorder on the atomic masses (\textit{i.e.} blurring the sublattice inversion), such as substitutional impurities, and hence a small coupling between both SPEs and TPL can exist. Additionally, atomic vacancies also break this sublattice symmetry. The vacancy defects can be deterministic created by means of irradiation\cite{fischer2021}, controlling the effective degree of dissipation, if needed.

For a general picture of the proposed mechanism, a system of $N$ SPEs interacting via a common -topologically-protected-phonon mode is considered in our theoretical model. To describe the TPL-mediated SPE-SPE interaction, we consider the common model of spin-boson interaction~\cite{Kopp_2007,Costi_2003}, where the SPE plays the role of a local spin $\frac{1}{2}$ system (\textit{i.e.} a two-levels system~\cite{Martinis_2002,Makhlin_2001}), with $|g\rangle =|\downarrow\rangle$ representing the ground state, and $|e\rangle = |\uparrow\rangle$ the excited state, respectively, in the $SU(2)$ eigenbasis of $\hat{\sigma}^{z}$. Hence the whole `spin-phonon' interacting system is governed by the following Hamiltonian (with $\hbar=1$)
\begin{align}\label{base}
    \mathcal{\hat{H}}
    =&
    \mathcal{\hat{H}}_{0}+\mathcal{\hat{H}}_{I}+\mathcal{\hat{H}}_{P},\\
    \mathcal{\hat{H}}_{0}
    =&
    \omega_{m}\hat{b}^{\dagger}\hat{b} + \sum_{j=1}^{N}\frac{\omega_{j}\hat{\sigma}^{z}_{j}}{2},\\
    \mathcal{\hat{H}}_{I}
    =&
    \sum_{j=1}^{N}g_{j}\hat{\sigma}^{z}_{j}\left(\hat{b}+\hat{b}^\dagger\right),\\
    \mathcal{\hat{H}}_{P}
    =&
    \sum_{j=1}^N \Omega_j\left(\hat{\sigma}^{+}_{j}e^{-i\omega_{P}t}+\hat{\sigma}^{-}_{j}e^{i\omega_{P}t}\right).
\end{align}
Here $\mathcal{\hat{H}}_{0}$ is the free Hamiltonian of the phonon and the SPEs, with energies $\omega_m$ and $\omega_j$, respectively, and $\hat{\sigma}^z$ is the third Pauli matrix. $\mathcal{\hat{H}}_{I}$ is the interaction Hamiltonian and $g_j$ is the spin-boson coupling strength (SPE-phonon in our system). Finally, $\mathcal{\hat{H}}_{P}$ represents an external pump provided by a laser of frequency $\omega_{P}$, in order to drive coherently both SPEs. The ladder operators are denoted $\hat{\sigma}^+, \hat{\sigma}^-$. The spin-boson Hamiltonian has been the subject of extensive studies in the context of strongly-correlated electronic systems, as an interesting scenario for the interplay between quantum criticality and entanglement. Former theoretical studies of this model\cite{Kopp_2007,Costi_2003,Martinis_2002,Makhlin_2001} have been concerned about the coupling between a two-level system and a dissipative phonon thermal bath. In this case, an exact mapping to the anisotropic Kondo model allows to show numerically (via NRG)\cite{Costi_2003} and analytically (via Bethe-Ansatz)\cite{Kopp_2007} that the corresponding spectral function of the phonon bath determines a crossover between an entangled regime and a disentangled one. In terms of this mapping, the corresponding Kondo temperature $T_K$ that determines the transition is proportional to the strength of the ``spin-flip operator''\cite{Kopp_2007}, a role that in our model is played by the external pump with magnitude $\Omega_j$.
We point out that, for some applications, it will not be necessary to consider the external pump, since the system will then evolve under the action of the Hamiltonian $\mathcal{\hat{H}}_{0}+\mathcal{\hat{H}}_{I}$ and some dissipative mechanism. More details are given when discussing particular features. A derivation of the atomic Hamiltonian -excluding $\hat{\mathcal{H}}_P$- is provided in the Appendix~\ref{sec:atomicHamiltonian}.

Further, by applying the unitary transformation $\mathcal{\hat{U}}=\exp{\left(i\sum_{j=1}^{N}\omega_{P}\hat{\sigma}^{z}_{j}t\right)}$ one gets the time-independent Hamiltonian
\begin{equation}
\begin{split}
    \mathcal{\hat{H}}' = &\,\omega_{m}\hat{b}^{\dagger}\hat{b}+\sum_{j=1}^{N}\left[\frac{\Delta_{j}\hat{\sigma}^{z}_{j}}{2}+g_{j}\hat{\sigma}^{z}_{j}\left(\hat{b}+\hat{b}^\dagger\right)\right.\\
    &\left.+\Omega_j\left(\hat{\sigma}^{+}_{j}+\hat{\sigma}^{-}_{j}\right)\right],    
\end{split}
\label{Hint}
\end{equation} 
with $\Delta_{j}=\omega_{j}-\omega_{P}$. 

To have a reasonable estimate of the order of magnitude of the parameters $\omega_m$ and $\omega_j$ in $\hat{\mathcal{H}}'$, we performed density functional theory (DFT) calculations, in close agreement with other approximations reported in the literature~\cite{jero216,Jiang2018}. Estimating the coupling strength $g_{j}$ from DFT is not straightforward and the details are  presented in Sec.~\ref{sec:atomicHamiltonian} of the Appendices.

\subsection{Master Equation}

To study the dynamics under realistic conditions, one should include the dissipation caused by the system-environment coupling. Consequently, we use the Markovian master equation (ME) for the system's density matrix ($\hat{\rho}$) to simulate the time evolution under the dissipative and decoherent mechanisms acting on the SPEs and phonon subsystems, respectively. The generalized ME is defined as follows
\begin{eqnarray}  \label{GME}
    \frac{d\hat{\rho}}{dt}=-\imath[\mathcal{\hat{H}}',\hat{\rho}]
    +\frac{\gamma_{b}}{2}\left(1+\Bar{n}_{b}\right)\mathcal{L}[\hat{b}]\hat{\rho}
    +\frac{\gamma_{b}}{2}\Bar{n}_{b}\mathcal{L}[\hat{b}^{\dagger}]\hat{\rho} \nonumber\\
    +\sum_{j=1}^N \frac{\gamma_{s}}{2}\left(1+\Bar{n}_{s}\right)\mathcal{L}[\hat{\sigma}^{-}_{j}]\hat{\rho}
    +\frac{\gamma_{s}}{2}\Bar{n}_{s}\mathcal{L}[\hat{\sigma}^{+}_{j}]\hat{\rho} +\frac{\gamma_{\phi}}{2}\mathcal{L}[\hat{\sigma}^{z}_{j}]\hat{\rho}
\end{eqnarray}
where $\Bar{n}_{b(s)}$ is the average number of quanta (\textit{i.e.} thermal phonons) in the reservoir corresponding to the phonon (SPEs) subsystem, respectively, and is calculated by the Bose-Einstein distribution as $\bar{n}_{b(s)}=\left(\exp{\left[\hbar\omega_{b(s)}/(\kappa_{B}T)\right]}-1\right)^{-1}$. Initially each subsystem is in thermal equilibrium with its reservoir at a given temperature $T$, and $\kappa_{B}$ is the Boltzmann's constant here. In a general case, the dissipation and decoherence mechanisms are characterized by $\gamma_{b}$ as the phonon damping rate, while $\gamma_{s}$ and $\gamma_{\phi}$ are the ``spin'' (\textit{i.e.} the local two-level system) rates for damping and pure dephasing, respectively. The Lindbladian superoperator is defined as $\mathcal{L}[\mathcal{\hat{O}}]\hat{\rho}=2\mathcal{\hat{O}}\hat{\rho} \mathcal{\hat{O}}^{\dagger}-\mathcal{\hat{O}}^{\dagger}\mathcal{\hat{O}}\hat{\rho}-\hat{\rho} \mathcal{\hat{O}}^{\dagger}\mathcal{\hat{O}}$.

\subsection{System initialization}
\textit{i) Initial phonon state.}
The phonon state is assumed to be initially in a thermal mixed state at temperature $T$, that in the coherent basis it can be written as
\begin{equation}\label{inphon}
    \hat{\rho}_{b}(0)=\frac{1}{\pi\bar{n}_{b}}\int|\xi\rangle\langle\xi|e^{-\frac{|\xi|^{2}}{\bar{n}_b}}d^{2}\xi,
\end{equation}
where $\xi$ in general is a complex number.

\textit{ii) Initial SPE state.} Let us consider that each SPE is prepared in a superposition state like $|\psi_{j}(0)\rangle =\alpha_j|g\rangle + \beta_j|e\rangle$, with $|g\rangle =|\downarrow\rangle$ representing the ground state, and $|e\rangle = |\uparrow\rangle$ the excited state, respectively, in the $SU(2)$ eigenbasis of $\hat{\sigma}^{z}$. Here, $|\alpha_j |^2 + |\beta_j |^2=1 $ and hence the density matrix for the subsystem of SPEs reads
\begin{equation}\label{inspin}
\hat{\rho}_s(0)=\otimes_{j=1}^N |\psi_{j}(0)\rangle \langle \psi_{j}(0)|.
\end{equation}
We point out that for $\alpha = \beta=1/\sqrt{2}$ one has the particular case of a qubit state, which we will consider for some further calculations as an optimal condition to observe interesting effects to be discussed.

\begin{figure*}[t] 
\centering
\includegraphics[width=0.34\textwidth]{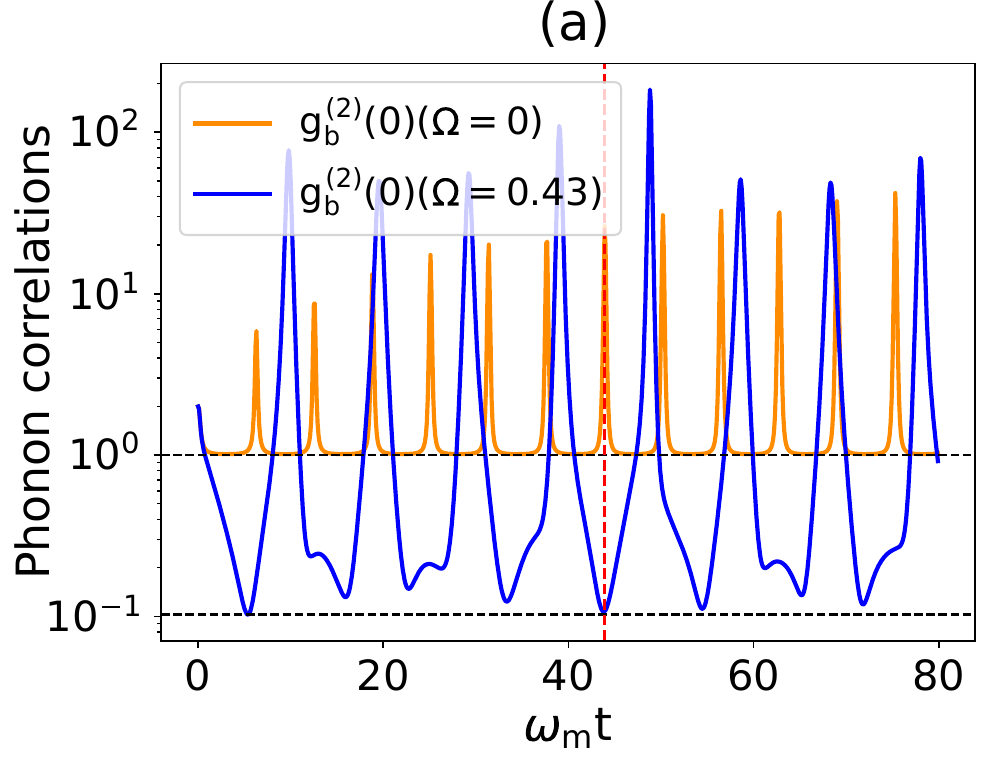}
\includegraphics[width=0.34\textwidth]{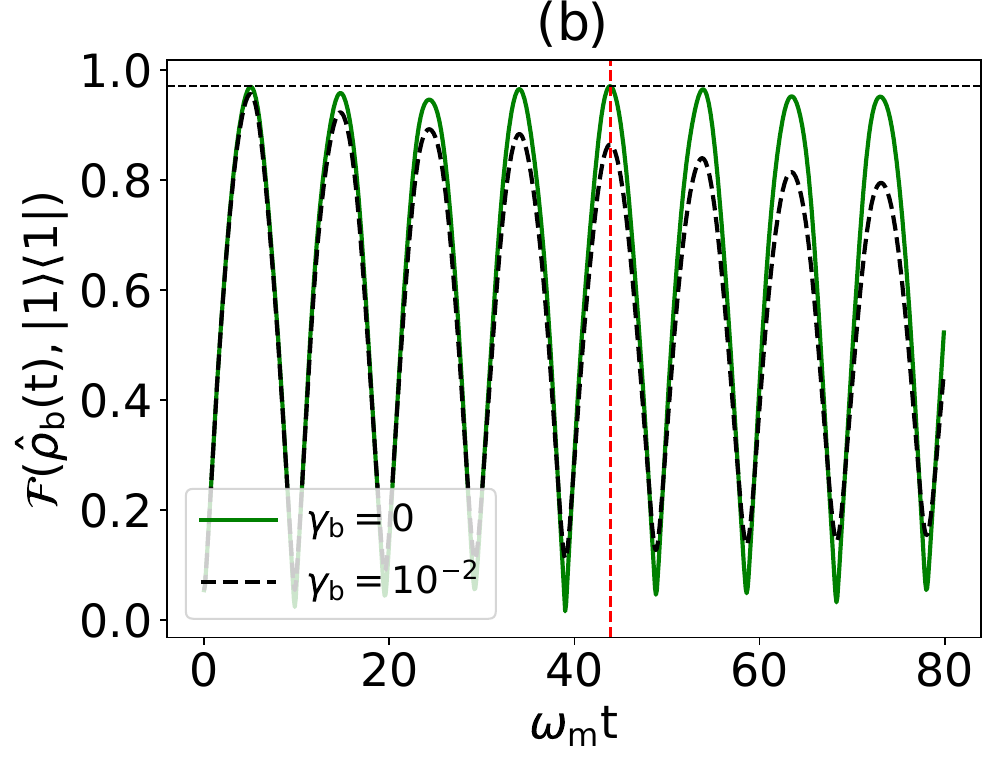}
\hspace{0.5cm}
\includegraphics[width=0.268\textwidth]{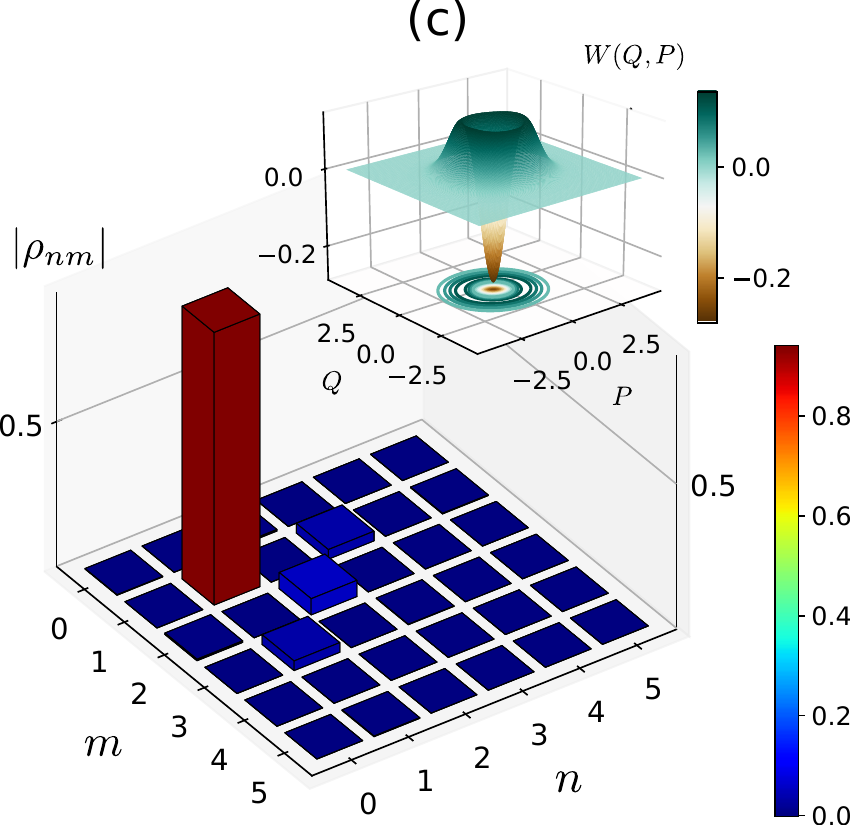}
\caption{Generation of single-phonon Fock state with one SPE initially in a superposition state with $\alpha=\beta=1/\sqrt{2}$ for $\Delta=0$, $\Omega=0.43$ and witnessed by: $(a)$ Second-order correlation function, $g^{(2)}_{b}(0)$. 
$(b)$ Fidelity of the Fock state $\vert 1 \rangle$ for different values of $\gamma_b$, i.e. TPL (green curve) and non-TPL (black curve).  $(c)$ Probability distribution in the basis of the Fock states at the dimensionless time $\omega_{m}t\approx43.9$, corresponding to the maximal fidelity $\mathcal{F}_{\vert 1 \rangle} \sim 0.97$. (inset: Negative values of Wigner function witnesses a quantum state). Other parameters in units of $\omega_m$ are $g_{1}=0.33$, $\gamma_{s}=\gamma_{\phi}=10^{-5}$ and $\bar{n}_{b}=0.003$, $\bar{n}_{s}=0$.}
\label{fig2}
\end{figure*}

\section{Results}
\label{sec:results}
\subsection{SPE-assisted generation of phonon quantum states}
\label{sec:ph-Fock}

\subsubsection{One SPE: Generation of phonon Fock states}
The purpose of this work is twofold, the first goal is to analyze an adequate physical model that effectively describes the SPE-phonon interaction within a TPL approach, and the second goal is to apply this model for the engineering of phonon quantum states and quantum correlations between SPEs, these being important resources for current quantum technologies. An important tool for controlling and transferring quantum information, for example in phononic based interfaces, is related to the creation of quantum states such as Fock, qubits, and Schr\"odinger cats (superposition of coherent states) \cite{Abdi2017,Sletten2019, Samanta2023,Bienfait2019, Montenegro2019}. In this context, in the following we propose such kind of quantum state engineering. Particularly, we show in a dissipative evolution the possibility of preparation of Fock and qubit states for the phonon mode in the TPL and non-TPL regimes, considering the loss mechanisms for SPEs. We find that phonon quantum states can in principle be generated for any number of SPEs. However, since the total dissipation in this dynamics (see Eq.~\ref{GME}) is proportional to the number of SPEs, then the corresponding fidelity of the resulting quantum states decreases with such multiplicity. Below we study the case of one SPE to fully analyze the observed effects, and in Appendix D we present the corresponding results for many SPEs. 

Let us consider the simplest case, when there is only one SPE interacting with the TPL and this SPE is driven by a coherent pump field, as defined in Eq.(4). In that particular case, the ME becomes
\begin{eqnarray}\label{1sME}
    \frac{d\hat{\rho}}{dt}=-\imath[\mathcal{\hat{H}}',\hat{\rho}]
    + \frac{\gamma_{s}}{2} \mathcal{L}[\hat{\sigma}^{-}]\hat{\rho}
     +\frac{\gamma_{\phi}}{2}\mathcal{L}[\hat{\sigma}^{z}]\hat{\rho},
\end{eqnarray}
where $\mathcal{\hat{H}}'$, as defined in Eq.~\ref{Hint}, is considered for $N = 1$ SPE in this case. As the phonon is topologically protected from the interaction with the environmental thermal phonons, \textit{i.e.} $\gamma_b=0$, then the terms related to the phonon damping in Eq.~\ref{GME} are absent here. Moreover, we emphasize that in our model the SPE energy corresponds to about 2~eV $\approx $ 483.6 THz, and hence the thermal environment at the room temperature gives $\bar{n}_s=0$, thus the ME (Eq.~\ref{GME}) reduces to Eq.~\ref{1sME}. 

Now let us assume that there is one SPE initialized in a superposition state (Eq.~\ref{inspin}) and driven by an external coherent source ($\hat{\mathcal{H}}_P$), while the phonons are initially in a thermal state (Eq.~\ref{inphon}) with $\bar{n}_{b}=0.003$. As a result of the SPE-TPL interaction and driving mechanism, Fock phonon states are generated at some intervals during the non-unitary dynamics, with the corresponding losses of the ``spin'' subsystem. To analyze the quantum statistics of the phonon dynamics, we numerically calculate the one-time second-order correlation function \cite{PRA100} for the phonon mode
\begin{align} \label{cohf}
    g^{(2)}_{b}(0)=\frac{\langle(\hat{b}^{\dagger})^2\hat{b}^2\rangle}{\langle\hat{b}^{\dagger}\hat{b}\rangle^2} 
\end{align}
Additionally, to witness the Fock state $\vert n \rangle$ we use complementary measures such as fidelity $\mathcal{F}=\sqrt{\langle n \vert \hat {\rho}_{b} \vert n \rangle }$ and the Wigner function. All these quantities are calculated in base of the reduced phonon state obtained as a partial trace on the ``spin'' subsystem of the global state by solving numerically \cite{Qutip} Eq.~\ref{1sME}, i.e. $\hat {\rho}_{b}= \text{Tr}_{s}[\hat {\rho}]$. 

\vspace{5pt}
\begin{figure*}[t]
\centering
\includegraphics[width=0.31\textwidth]{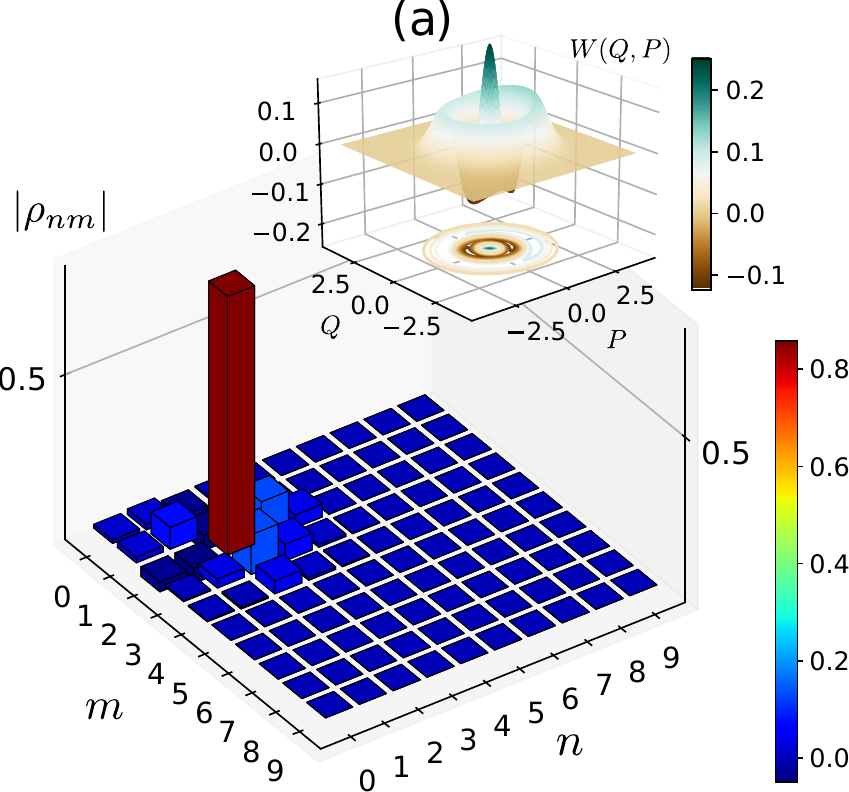}
\includegraphics[width=0.31\textwidth]{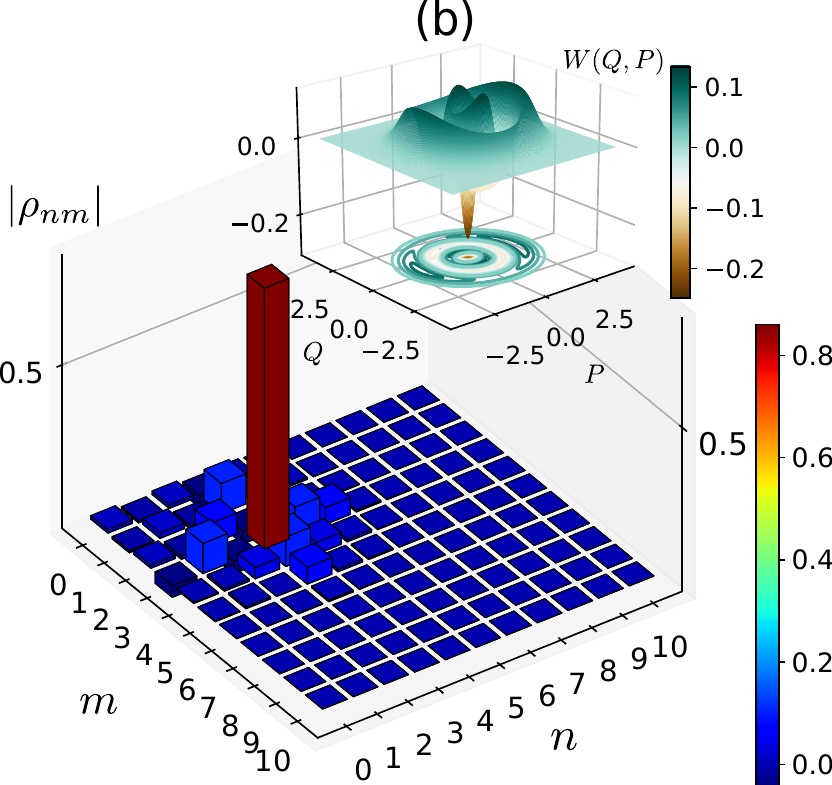}
\includegraphics[width=0.31\textwidth]{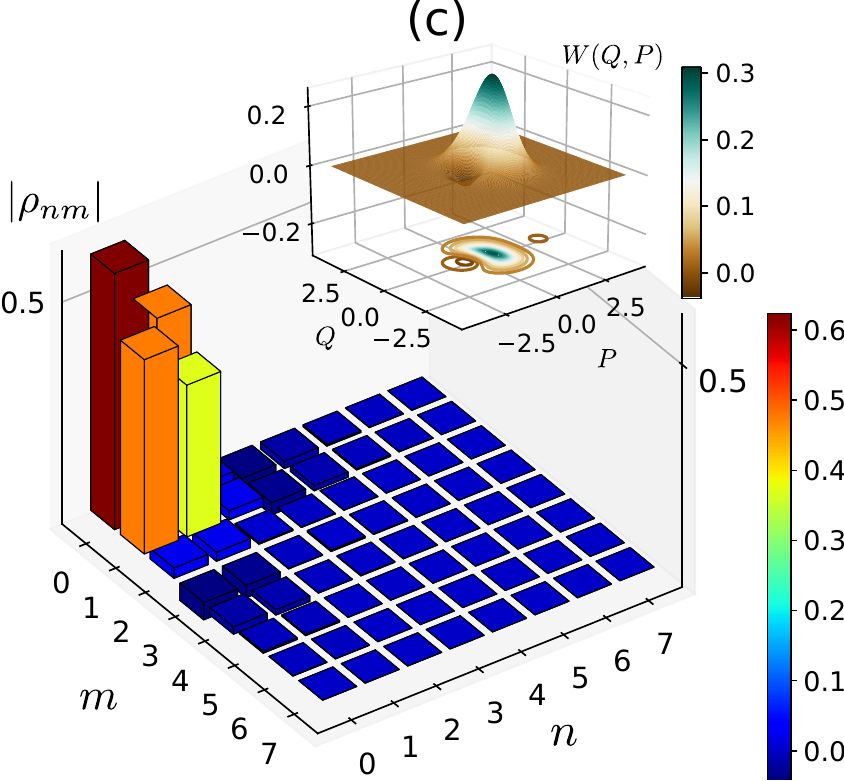}
\caption{Dynamical preparation of the phonon Fock and Qubit states with one SPE: $(a)$ Fock state $\vert n=2 \rangle$ prepared for $\alpha=0.30$, $\Omega=0.63$ and $\Delta=1.23$. The maximal fidelity $\mathcal{F}_{\vert 2 \rangle}=0.92$ is observed at $\omega_{m}t=
146.5$. $(b)$ Fock state $\vert n=3 \rangle$ prepared for $\alpha=0.55$, $\Omega=1.30$ and $\Delta=0.60$, with the maximal fidelity $\mathcal{F}_{\vert 3 \rangle}=0.92$ occurring at $\omega_{m}t=344.6$. $(c)$ Phonon Qubit state ($\vert 0 \rangle + \vert 1 \rangle)/\sqrt{2}$ with maximal fidelity $\mathcal{F}_{qubit}=0.98$ at $\omega_{m}t=66.2$ for $\alpha=1$, $\Omega=0.27$ and $\Delta=1.87$. Other parameters are the same as in Fig.~\ref{fig2}. }
\label{fig3}
\end{figure*}
\textit{Single-phonon Fock state.} 

The methodologies and setups for the preparation of individual phonon states \cite{Cohen2015, Sollner2016, Osada2022} have attracted great interest in the last decade, due to the many practical applications in quantum technologies. In the following we discuss the effect of the single-phonon Fock state within our proposal. 
In Fig.~\ref{fig2} we clearly see that, at some instants of dimensionless time, a phonon Fock state $\vert n=1 \rangle$ emerges. This effect is consistent with the following: \textit{(i)} $g^{(2)}_{b}(0)<0.5$ (see Fig.~\ref{fig2}a); and \textit{(ii)} the probability distribution on the Fock basis, and a negative value of the Wigner function (see Fig.~\ref{fig2}c and inset). 
We solved the ME as a function of the parameters $\{\Omega,\Delta,\alpha\}$, and we found that the single-phonon Fock state is generated with the fidelity $\mathcal{F}_{\vert 1 \rangle} \sim 0.97$ at the dimensionless time $\omega_{m}t=43.9$, and it is almost preserved periodically in time for the case of the TPL, as shown by the green solid line in Fig.~\ref{fig2}b. In contrast, in a regular phononic waveguide with losses (\textit{i.e.} non-topologically protected) the fidelity of the Fock state decreases substantially in time, as shown by the black-dashed curve in Fig.~\ref{fig2}b. This quantum effect for the TPL indicates that a coherently driven SPE, during its dynamics, exhibits time intervals when the generation of single phonon states at high fidelity is stimulated - a great practical application in many fields. 

\vspace{5pt}
\textit{Fock states $\vert n=2 \rangle$ and $\vert n=3 \rangle$.} 

It is possible to prepare other Fock states, e.g. $\vert n=2 \rangle$ and $\vert n=3 \rangle$, by appropriately tuning the parameters $\Omega$, $\Delta$ and $\alpha$. For instance, the Fock state $\vert 2 \rangle$ is optimally prepared for the combination of parameters presented in Fig.~\ref{fig3}a. This effect can be witnessed by the correlation function $g^{(2)}_{b}(0)\to 0.5$ (not presented here), and further corroborated by the probability distribution in the Fock basis, as well as by the negative value of the Wigner function, thus indicating the quantum state, see inset in Fig.~\ref{fig3}a. The fidelity of this state reaches $ \mathcal{F}_{\vert 2 \rangle} \sim 0.92$. On the other hand, the Fock state $\vert n=3 \rangle$ can be prepared in a similar way, see Fig.~\ref{fig3}b. For the sake of space, we do not show explicitly the figures containing the dynamics of the second order correlation function and the fidelities for the Fock states considered here, however these quantities were also analyzed and they display similar results as the ones shown in Figs.~\ref {fig2}a-b. Based on these results, the information of the time instants corresponding to the maximum values of the fidelities is obtained and indicated in Fig.~\ref{fig3}.

\subsubsection{One SPE: Generation of the phonon Qubit state}

Since the phonon Fock states are realizable in our proposal, one can expect to prepare a phonon qubit, which is in fact a superposition of the Fock states $\vert 0 \rangle$ and $\vert 1 \rangle$. A phonon qubit state can serve as a key element in a solid-state quantum network, thus representing a candidate to replace its ``cousin'', the photonic qubit, used in photon-based setups \cite{Niemietz2021}. Methodologies for preparing phonon qubit states nowadays are limited to very few types of interfaces, particularly based on micromechanical resonators \cite{Reed2017, Wollack2022, Samanta2023}. Here, we show that a high-fidelity phonon qubit, $(\vert 0 \rangle + \vert 1 \rangle)/\sqrt{2}$, could be prepared in our proposal using the TPL configuration. In Fig.~\ref{fig3}c one sees clearly how the qubit state is witnessed by the probability distribution in the Fock basis and the negative part of the Wigner function (see inset). 

\subsection{Phonon-assisted SPE-SPE Quantum Correlations}
\subsubsection{Generation of transient maximally-entangled states in a dissipative dynamics}

In this subsection and the next one, we are interested in studying the quantum correlations established in time between two distant SPEs that interact with a common phononic mode, be it TPL or non-TPL. The quantum correlations will be measured using the Concurrence to witness the Entanglement. 
\begin{figure}[t]
\centering
\includegraphics[width=0.4\textwidth]{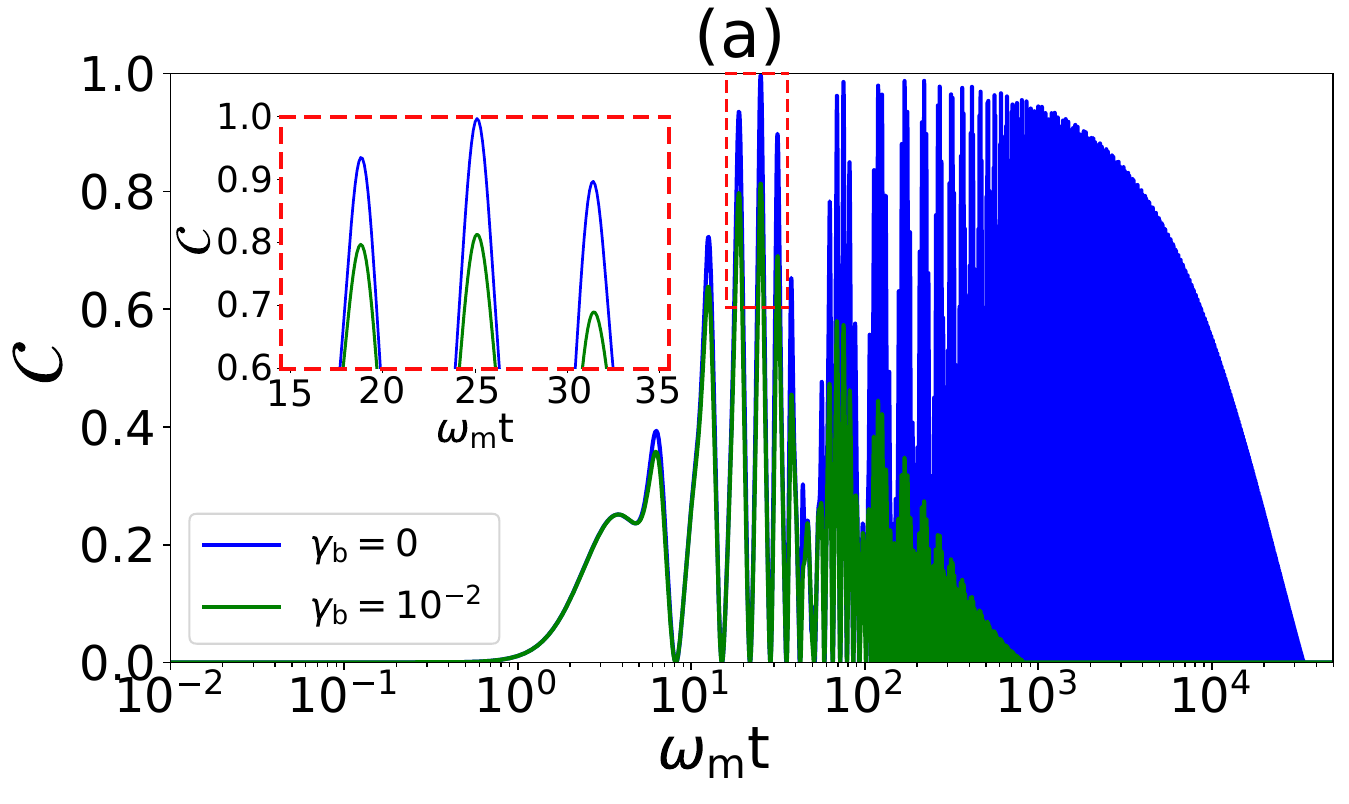}
\includegraphics[width=0.24\textwidth]{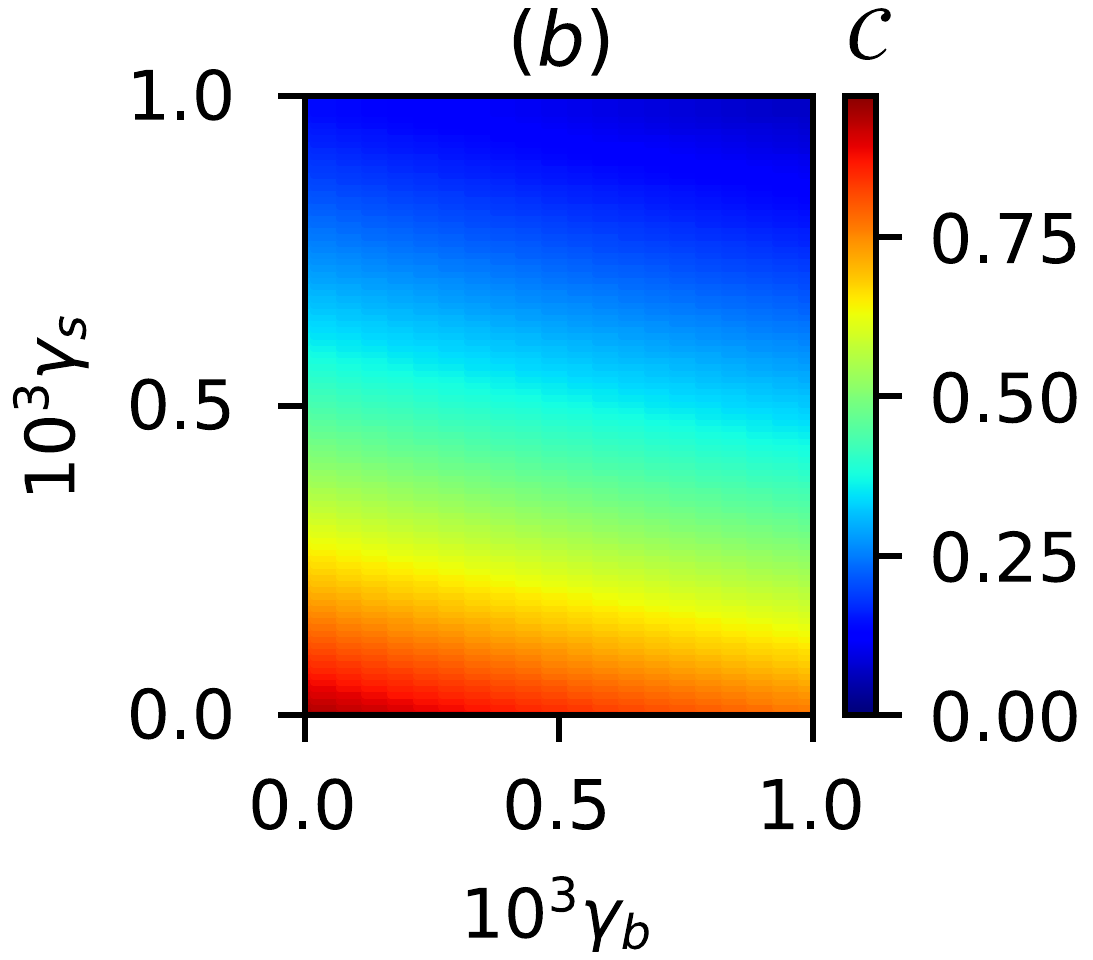}
\caption{Dynamics of Concurrence ($\mathcal{C}$) for two SPEs initially in a superposition state (Eq.~\ref{inspin}) with $\alpha = \beta=1/\sqrt{2}$. $(a)$ Concurrence for TPL (blue) and non-topological phonon (green) calculated for SPE losses as $\gamma_{s}=\gamma_{\phi}=10^{-5}$; (inset: zooming of a MES for TPL). $(b)$ Mapping the Concurrence at the dimensionless time $\omega_{m}t \approx 25$ as function of the SPE and phonon damping rates. Other parameters in units of $\omega_{m}$ are: $g_{1}=g_{2}=0.33$, $\Omega=0$, $\bar{n}_{b}=0.003$, $\bar{n}_{s}=0$.
}
\label{fig4}
\end{figure}

Let's consider two SPEs initialized in the state defined by Eq.~\ref{inspin}, and as compared to the previous protocol of generation of phonon Fock states, here is not necessary to include a driving field, \textit{i.e.} $\Omega=0$ in Eq. \ref{Hint}. On the one hand, we take into account the decoherence mechanisms for SPEs such as damping and dephasing and, on the other hand, the TPL is protected from decoherence. In order to compare with a non-topologically protected phonon mode, we calculate the quantum correlations considering the phonon damping rate $\gamma_b\neq0$ within the bath at the temperature $T$. Therefore, the ME (Eq.~\ref{GME}) will be numerically solved for these two configurations.

The quantum correlations witnessed here as Entanglement for a system of two qubits (SPEs) can be quantified by the Concurrence ($\mathcal{C}$) \cite{Wootters}.  
We resume here the conceptual calculation of the Concurrence. For a general mixed state $\hat{\rho }_{AB}$ of two qubits, one defines $\widetilde{\hat{\rho} }$ to be the ``spin''-flipped state $\widetilde{\hat{\rho}}_{AB}=(\hat{\sigma}^{y}\otimes \hat{\sigma}^{y})\hat{\rho}_{AB}^{\ast }(\hat{\sigma}^{y}\otimes \hat{\sigma}^{y})$, where $\hat{\rho}^{\ast}$ is the complex conjugate of $\hat{\rho}$, and $\hat{\sigma}^{y}$ is the Pauli operator. Therefore, the Concurrence is defined as 
\begin{equation} \label{Conc}
\mathcal{C}(\rho )=\text{max}\{0,\lambda _{1}-\lambda _{2}-\lambda _{3}-\lambda_{4}\},
\end{equation}
where $\{\lambda_{i}\}$ are the square roots in decreasing order of the eigenvalues of the non-hermitian matrix $\hat{\rho}\widetilde{\hat{\rho}}$. To point out here, the Concurrence normalized to one witnesses an entangled two-qubit state $\hat{\rho}_{AB}$ by obeying the condition $0<\mathcal{C}\leq 1$. Hence, one has a maximally entangled state (MES) when $\mathcal{C}=1$, and for $\mathcal{C}=0$ there is no entanglement at all, so the two-qubit state can be written as a product of separated states, i.e. $\hat{\rho}_{AB}=\sum p_i \hat{\rho}^A_i \otimes \hat{\rho}^B_i$. Famous examples of MES are the well-known four Bell states \cite{Horodecki4}, however, it is possible to have other types of MES, which are not necessarily Bell states.

In the following, we study the evolution of the Entanglement between the two SPEs. For example, in Fig. \ref{fig4}a, extensive time dynamics of Entanglement ($\mathcal{C}$) are shown for the TPL (blue curve) and for phonon damping (green curve) models. Both correlations are promptly generated, and at an early stage exhibiting a MES ($\mathcal{C} \sim 1$) for the topologically protected configuration, see the inset in Fig. \ref{fig4}a (blue curve). The MES is very useful for quantum information applications \cite{Bennett2000, Horodecki4,PRA2016}. However, the quantum correlations decrease exponentially under the dissipation mechanisms of the SPEs and are totally lost at long times. The quantitative effect of such dissipation on the Entanglement is shown in Fig. \ref{fig4}b. One sees that the Entanglement is a very fragile quantum resource that can be quickly destroyed by strong dissipation, however it is possible to engineer some protection protocols like the one proposed here, via topological phonon state. 

An important remark is that in our model the distance between the SPEs can be arbitrarily large without compromising the coherence. In general, there are two mechanisms of decoherence related with distances:
\begin{itemize}
    \item In electronic systems such as the SPEs, the interactions responsible for the coherence decrease exponentially with distance (\textit{e.g.} a direct SPE-SPE interaction should suffer from this).
    \item The number of particles to interact will increase with distance, thus enhancing the phonon damping. In one-dimension, the damping should increase geometrically (\textit{e.g.} a non-topologically protected phonon line should be affected).
\end {itemize}
As a main result of this subsection, we show in Fig. \ref{fig4} the spontaneous generation of SPE-SPE quantum Entanglement (inseparable bipartite state) only stimulated by the phonon-mediated interaction, since in this protocol there is no external driving of SPEs, in contrast to the protocol of phonon Fock state generation (sec. \ref{sec:ph-Fock}). However, such Entanglement generation protocol is not stable in time, particularly here is rapidly oscillating and can be not very suitable for practical purposes. On the other hand, to achieve a stable Entanglement over a long time period one can look for a steady-state in the open system. We consider such mechanism in the next subsection, where a stable Entanglement is indeed obtained.

\begin{figure*}[t]
\centering
\includegraphics[width=0.31\linewidth]{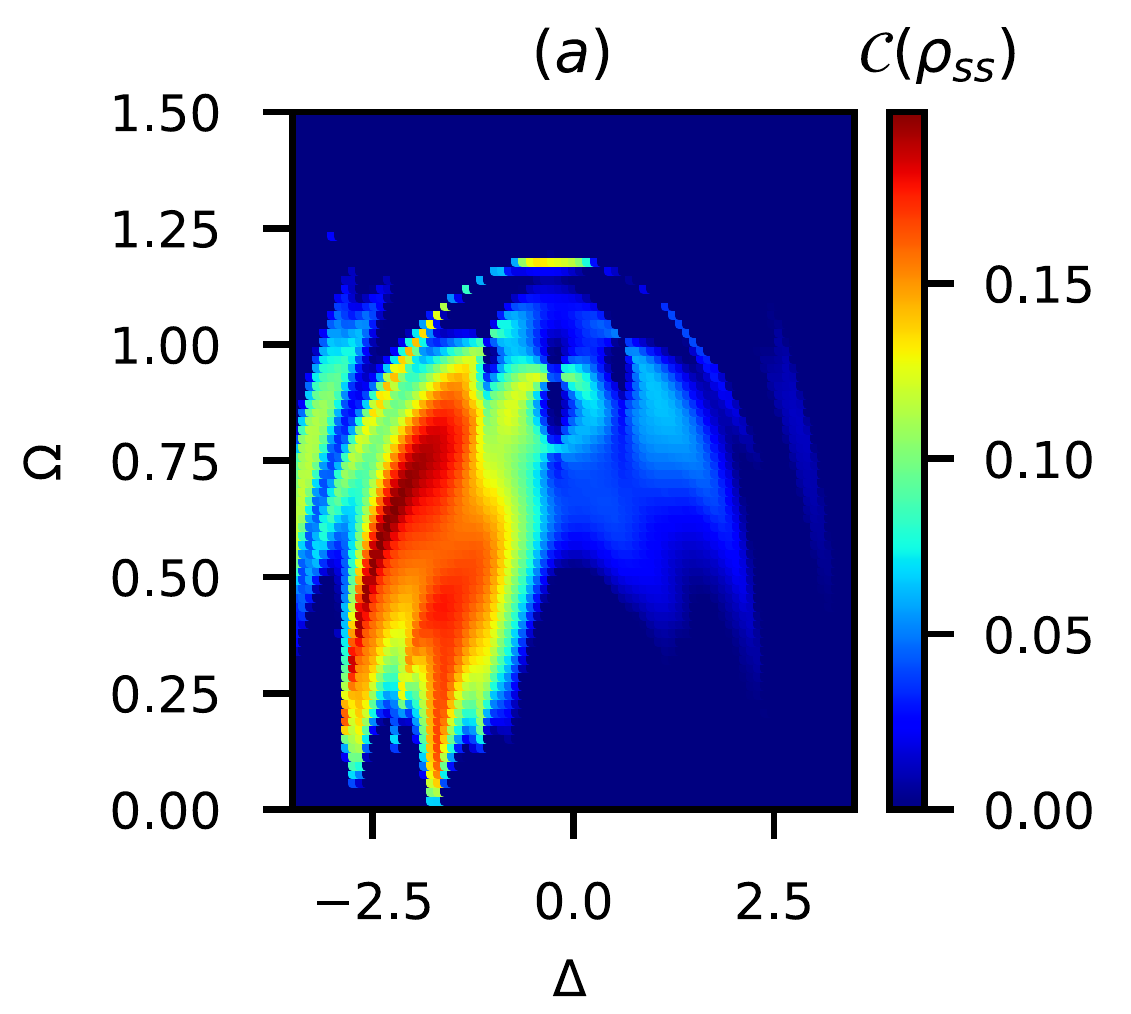}
\includegraphics[width=0.3\linewidth]{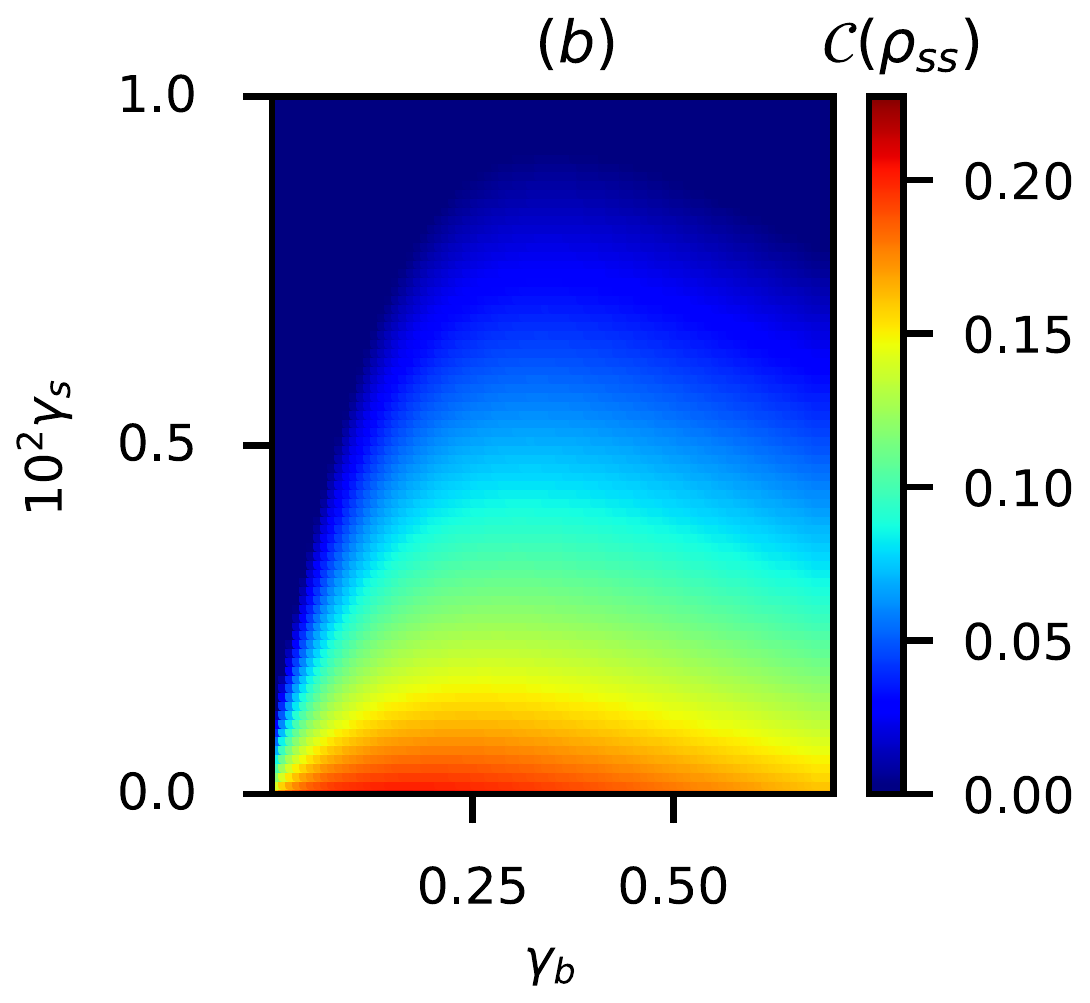}
\includegraphics[width=0.6\linewidth]{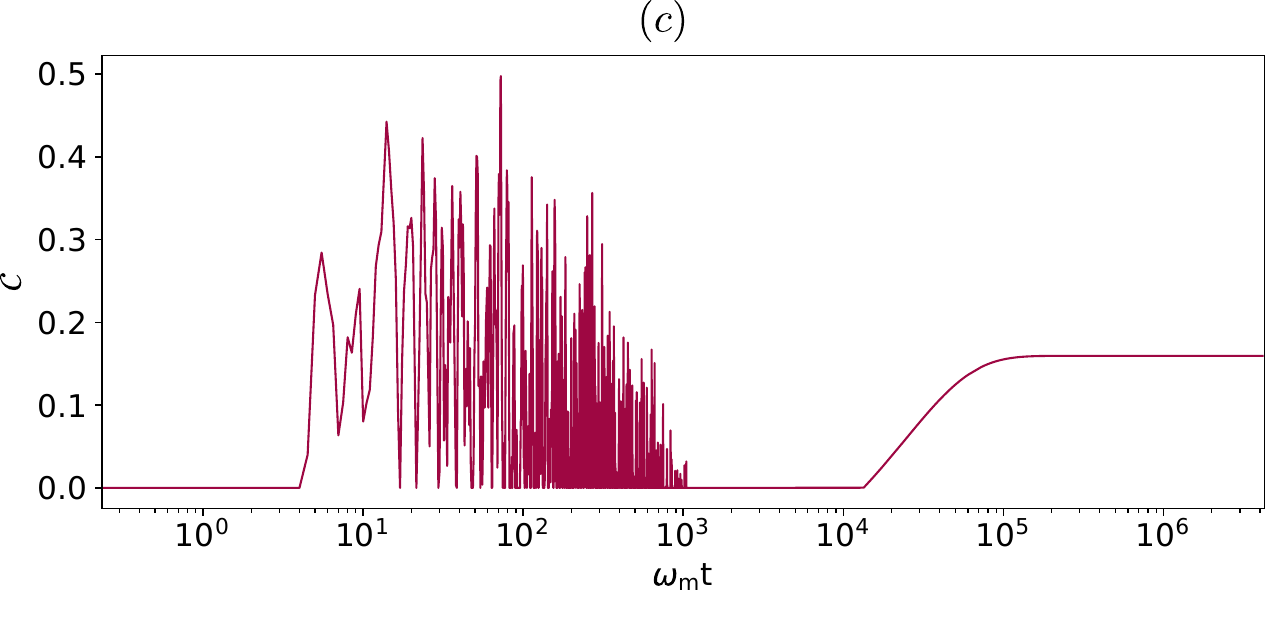}
\caption{The heatmap for steady-state Concurrence, $\mathcal{C}(\rho_{ss})$, as a function of parameters: $(a)$ Laser driving ($\Omega$) and spin-laser detuning ($\Delta$), for the spin decay rates $\gamma_{s}=\gamma_{\phi}=10^{-5}$ and phonon decay $\gamma_{b}=10^{-3}$; $(b)$ Spin decays (where $\gamma_\phi=\gamma_s$) and phonon decay rate ($\gamma_b$), for the parameters $\Omega=0.61$ and $\Delta=-0.98$, chosen from the map $(a)$ to have a higher Concurrence. Here the spin-phonon coupling $g_1=g_2=0.33$, $\bar{n}_{s}=0$ and $\bar{n}_{b}=0.003$. $(c)$ Driven-dissipative dynamics of the Entanglement quantified by Concurrence for two SPEs initialized in a ground state for the following parameters in units of $\omega_m$: $\Omega=0.61$, $\Delta=-0.98$, spin decays $\gamma_{s}=\gamma_{\phi}=10^{-5}$, and phonon decay $\gamma_{b}=10^{-3}$. Here the spin-phonon coupling $g_1=g_2=0.33$, $\bar{n}_{s}=0$ and $\bar{n}_{b}=0.003$, i.e. similar as in Fig.2.
}
\label{fig5}
\end{figure*}

\subsubsection{Generation of steady state Entanglement in a driven-dissipative dynamics}

In the following we are interested to study the possibility to entangle the SPEs for a long time period, therefore a steady state solution of the ME~\eqref{GME} can provide stable quantum correlations in the limit of $t \to \infty $. However, for a steady state solution, a driving source is required to balance the damping terms in the ME describing the open quantum system. Hence, let us consider the generalized ME~\eqref{GME} where the dissipation for spins and phonon are considered, and additionally the spins are driven by an external field represented by the Rabi frequency $\Omega$ in the Hamiltonian~\eqref{Hint}. With this in mind, one has the following ME equation
\begin{eqnarray}
\label{ME2}
    \frac{d\hat{\rho}}{dt}&=&-\imath[\mathcal{\hat{H}}',\hat{\rho}] + \frac{\gamma_{b}}{2}\left(1+\Bar{n}_{b}\right)\mathcal{L}[\hat{b}]\hat{\rho}
    +\frac{\gamma_{b}}{2}\Bar{n}_{b}\mathcal{L}[\hat{b}^{\dagger}]\hat{\rho} \nonumber \\
    &+& \sum_{j=1}^2 \frac{\gamma_{s}}{2} \mathcal{L}[\hat{\sigma}_j^{-}]\hat{\rho}
     +\frac{\gamma_{\phi}}{2}\mathcal{L}[\hat{\sigma}_j^{z}]\hat{\rho},
\end{eqnarray}
where $\mathcal{\hat{H}}'$ is defined in Eq.~\eqref{Hint}, and considering $\bar{n}_s=0$ as explained in the text following the Eq.~\eqref{1sME}. 
\begin{figure*}[t]
\centering
\includegraphics[width=0.495\linewidth]{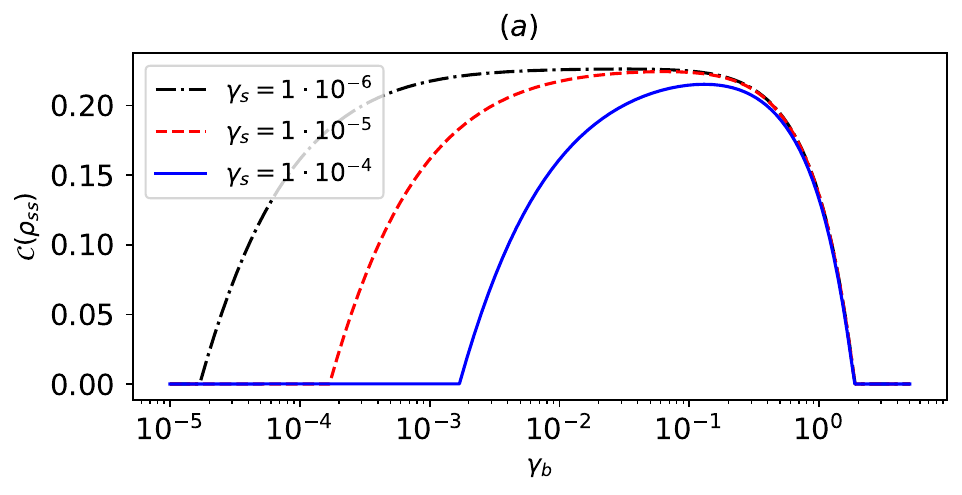}
\includegraphics[width=0.49\linewidth]{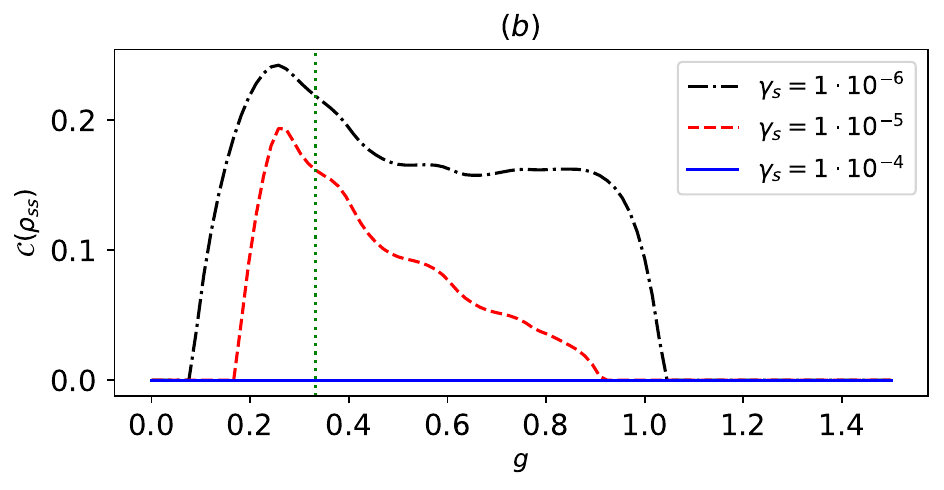}
\caption{Steady-state Concurrence, $\mathcal{C}(\rho_{ss})$, for fixed $\Omega=0.61$ and $\Delta=-0.98$ and varying: $(a)$ the decays of spins and phonon; $(b)$ spin-phonon coupling ($g$) and decays of spins so that $\gamma_s=\gamma_\phi$, here $\gamma_b=10^{-3}$. The green dotted vertical line corresponds to $g=0.33$, resulted from the microscopic calculations (Appx. \ref{sec:atomicHamiltonian}). Other parameters are the same as in previous figure.}
\label{fig6}
\end{figure*}


To achieve the steady state Entanglement, we solve by an exact numerical method \cite{Qutip} the equation $d\hat{\rho}/dt=0$, and with the steady state solution $\rho_{ss}$ we compute the concurrence $\mathcal{C}(\rho_{ss})$ using the same procedure as in Eq.~\eqref{Conc}. In Fig. \ref{fig5}(c) we show the dynamics of the concurrence, where at a late time ($\omega t \sim 10^5$) the Entanglement is stabilized in a stationary state, so that $\mathcal{C}(\rho_{ss}) \approx 0.16$. Despite the steady state here does not exhibit a maximal Entanglement, for practical purposes a robust partial Entanglement in such a solid state system could be an important resource to be used for distant quantum communication by applying the protocols of entanglement distillation (purification), in order to produce high-fidelity maximally entangled pairs of qubits even in noisy environment \cite{BBPSSW, EntDist,nature2006}. We present more details about such possibilities in Sec. ~\ref{sec:disc}. To analyze the steady Entanglement in a more detailed manner on the system's parameters, we present a comprehensive study represented in Figs.~\ref{fig5}-\ref{fig6}. For example, in Fig.~\ref{fig5}(a) the mapping of the steady value of the Concurrence $\mathcal{C}(\rho_{ss})$ in the space of the parameters $\Omega$ and $\Delta$ is shown. We observe that $\mathcal{C}(\rho_{ss})>0$ occurs especially for the negative values of the spin-laser detuning, and the optimal Entanglement is concentrated in the orange-brown color region. As a result of this map, one can choose the optimal combinations of $\Omega$ and $\Delta$. A similar concept is used to map the Concurrence in the space of the damping parameters as the spin decay rates, considering $\gamma_\phi=\gamma_s$, and phonon decay rate $\gamma_b$, see Figs.~\ref{fig5}(b) and \ref{fig6}(a). Another important parameter in our model is the spin-phonon coupling $g$, and therefore in Fig.~\ref{fig6}(b) one may identify quantitatively the region of the  coupling where optimal steady Entanglement is produced. The value of the spin-phonon coupling resulting from the microscopic calculation (Appendix \ref{sec:atomicHamiltonian}) fits well in the region of strong Entanglement, see the green dotted vertical line in Fig.~\ref{fig6}(b). Based on this quantitative analysis, i.e. figures \ref{fig5}-\ref{fig6}, the regions of optimal parameters can be defined to achieve strong steady Entanglement. The identification of these optimal parameters provides great support for the eventual experimental studies.

\section{Discussion} 
\label{sec:disc}

For the proposed physical configuration, i.e. one or more optically active defects (SPEs) in hBN coupled by a grain boundary that acts as a waveguide for topologically protected phonon modes, we have studied the possibility of generating quantum correlations. We based our analysis on the spin-boson model, to represent the coupling between the two-level electronic degrees of freedom at the defect with the topological phonon modes, and we included
an external laser pumping into the system dynamics.  In this scenario, we identified several configurations in the parameter space which are, in principle, suitable to sustain long-term quantum correlations, thus representing attractive possibilities for quantum information technologies. Our results indicate that, to sustain long-term correlations, it is essential for the system to be open, such that it is subject to a steady balance between dissipation and external laser pumping.
We applied the Concurrence as an estimator for such correlations, but of course there exists other complementary definitions, particularly the Quantum Discord ($\mathcal{QD}$). In Appendix~\ref{appQDNew}, we analyzed quantitatively the evolution of the $\mathcal{QD}$, and we found a consistent behaviour as the one reported for the Concurrence (see Fig.~\ref{fig9}). Therefore, our analysis remains valid for different estimators of quantum correlations.

In what follows, we shall test our theoretical results by comparison with an experimental-like simulation on a Quantum IBM for the system dynamics and correlations, showing an excellent agreement between them. Finally, we shall also discuss on the possible mechanisms for improving the Entanglement fidelity, of practical importance for potential applications in quantum information technologies.

\subsection*{\textit{Experimental simulation} of the Quantum Correlations on Quantum IBM } 
\label{sec_exp}
\begin{figure}[b]
\centering
\includegraphics[width=0.98\linewidth]{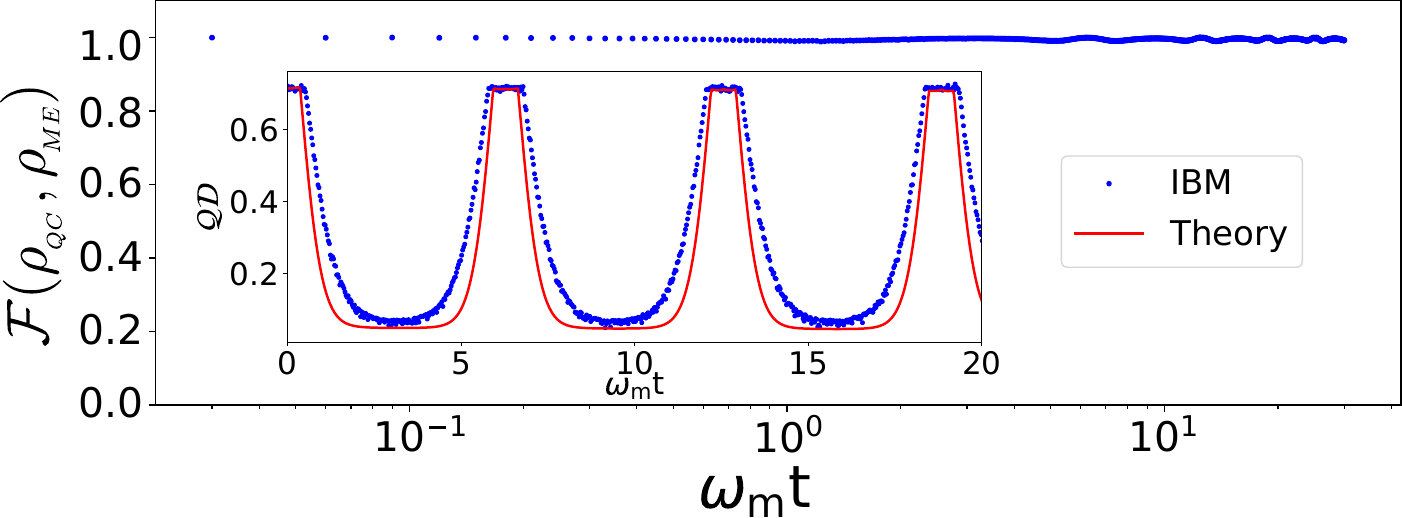}
\caption{Fidelity of the evolved state on quantum computer ($\rho_{QC}$) compared to the state ($\rho_{ME}$), calculated theoretically by Eq.~\ref{ME2}. Inset: Quantum Discord ($\mathcal{QD}$) for two SPEs initialized in a Bell-diagonal state (see Appx. \ref{appQD} Eq.~\ref{BDS}). The solid lines are the solutions on a classical device (common computer), the dots are generated using the IBM Quantum Qiskit simulator. The parameters here are the same as in Fig. \ref{fig9}(b), \textit{i.e.} for TPL ($\gamma_b=0$).}
\label{fig_IBM}
\end{figure}
In the following, we propose to simulate an experimental measurement of the evolution of the quantum correlations in case of the SPEs initialized in the Bell-diagonal state as in Eq.~\ref{BDS}. For this purpose, we use the IBM Qiskit simulator \cite{IBM}, where the intrinsic quantum noise is taken into account during the calculation, thus simulating a quasi-realistic quantum measurement. This quantum experiment is emulated via a quantum circuit, which is presented in details in the Appendix \ref{appIBM}. 

Solving the Lindblad ME, see Eq.~\ref{ME2}, on a classical computer and on the IBM’s quantum simulator, the quantum correlations quantified by Quantum Discord (see Appendix \ref{appQDNew}) are shown in Fig. \ref{fig_IBM}. The solid lines represent the solution of the master equation on a classical computer, and the dots represent the result from the quantum simulator. A very good correspondence is found between the theoretical and quasi-experimental results, thus demonstrating the possibilities of existing quantum simulators and processors such as IBM, to be successfully used to emulate results similar to the experimental ones in problems like the one proposed in this work.

\subsection*{Mechanisms of improving the Entanglement fidelity}

Quantum correlations are vulnerable to unavoidable errors by decoherence and losses, and our proposed configuration is not an exception. Nevertheless, protocols for improving the quality (fidelity) of partial entanglement are available, particularly \textit{Entanglement distillation (ED) or purification} \cite{BBPSSW, EntDist,nature2006}. For instance, the experimental protocol for ED proposed by Kalb, \textit{et al}.~\cite{EntDist} for a solid-state quantum network, with nodes containing NV-centers in diamond as a communication qubit and a C-13 nuclear spin as a memory qubit, would be suitable to our proposed system, due to the similarities between quantum states in NV centers and those in defects in 2D materials.

\section{Methods}
\label{sec:methods}

The electronic structure calculations were made with density functional theory, implemented by the VASP code\cite{vasp1,vasp2,vasp3}. We used a cutoff of 500 eV, PAW\cite{paw} pseudopotentials and the Perdew-Burke-Ernzerhof decription of the exchange-correlation\cite{pbe}. A single k-point ($\Gamma$) was used in the supercell calculation. Pyprocar\cite{pyprocar} was used for the analysis of the electronic structure, and Phonopy\cite{phonopy2} for the calculation of phonons. The photoluminiscence spectra was calculated by following Ref~\cite{Alkauskas_2014}. Some preliminary calculations were made with density functional tight-binding\cite{hourahine2020dftb,mio}. For the numerical solving of the master equations and calculations of the quantum states and correlations we use the Quantum Toolbox in Python - "QuTiP" \cite{Qutip}. The quantum circuit was emulated on the IBM Qiskit simulator \cite{IBM}.

\section{Conclusions}
\label{sec:concl}

The two-dimensional material hexagonal boron nitride (hBN) has two types of defects with great potential for applications in quantum information technologies: single photon emitters (SPEs) and topologically protected phonon lines (TPL). While the first are localized defects, the second represent a line of atomic defects (\textit{e.g.} a grain boundary) with almost dissipationless phonon states. As the typical phonon frequencies of both defects are similar, a sizable interaction develops among them. This allows the use of the TPL as a topologically protected one-dimensional waveguide, connecting distant SPEs within a single layer of hBN. Upon laser pumping, this waveguide preserves quantum correlations (characterized by Entanglement and Quantum Discord) for comparatively long periods of time, thus allowing for quantum information encoding and transmission.

This proposal shed light on the possibilities offered by this material-built-in nano-architecture as a phononic quantum device. By means of a simple model, we showed that in a dissipative system -upon laser pumping- one SPE can excite a single as well as larger Fock states (\textit{e.g.} $\vert n=2 \rangle$ and $\vert n=3 \rangle$) of the TPL. In addition to the Fock states, we showed the possibility to create a phonon qubit state with a high fidelity, $\mathcal{F} \sim 0.98$. Remarkably, when two SPEs interact with the TPL -plus the pumping field- the Fock and qubit states can also be excited, but at lower fidelities because the global losses increase. In this way, two SPEs can interact via the TPL by exchanging phonons. We showed that both SPEs remain entangled for several cycles of the evolution, as measured by the quantum concurrence. Remarkably, the coherence is practically independent of the spatial distance between both SPEs, thanks to the topological protection against scattering provided by the TPL. 

In summary, our theoretical results provide a foundation to consider arrays of SPEs connected by TPL phononic waveguides as promising candidates for quantum information technologies. Our approach also opens a new direction to explore more complex models, accounting for additional factors, thus enriching our understanding of this system and its applications.

\vspace{5pt}


\begin{acknowledgments}
This research was funded by FONDECYT through grants 1231487, 1220715, and 1230440; by the Center for the Development of Nanosciences and Nanotechnology, CEDENNA AFB 220001; Conicyt PIA/Anillo ACT192023. Powered@NLHPC: This research was partially supported by the supercomputing infrastructure of the NLHPC (ECM-02).
\end{acknowledgments}

\appendix

\section{Derivation of the microscopic Hamiltonian}
\label{sec:atomicHamiltonian}

\begin{figure*}[ht]
    \centering
    \includegraphics[width=\textwidth]{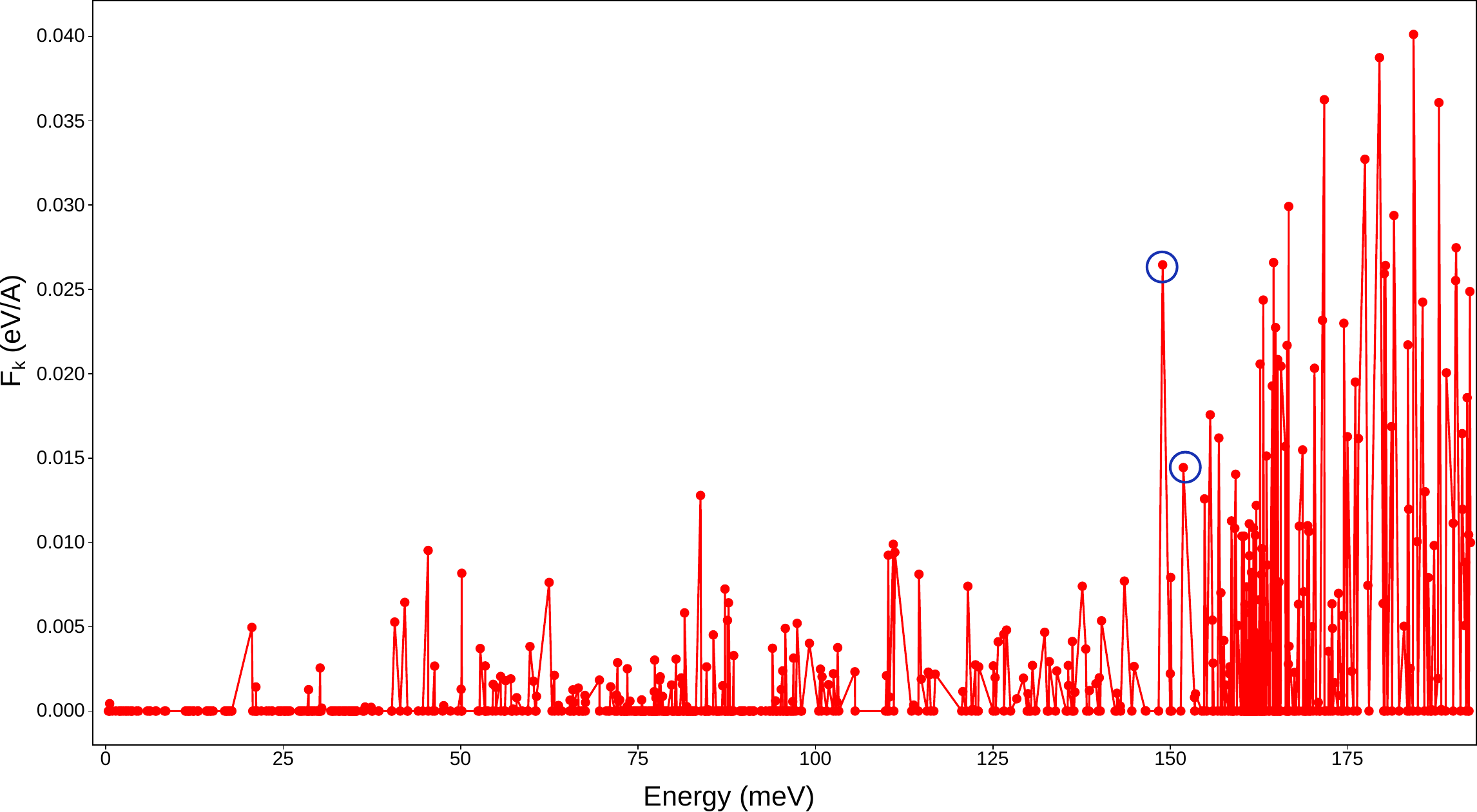}
    \caption{Forces of the excited state in the ground state geometry, $F_k$, as a function of the phonon mode. The forces already are in the basis of the phonons. The modes belonging to the topologically-protected line are in a black circle. }
    \label{fig:Fk}
\end{figure*}
Let's consider a SPE with only two states involved in the transition. The energy depends parametrically on the nuclear coordinates:
\begin{equation}
H(\mathbf{R}) = \sum_{i=1}^2 E_i(\mathbf{R})c_i^\dagger c_i,
\end{equation}
\noindent where $E_i$ refers to the energy of each level, $c_i$ is a fermionic annihilation operator,  $i$ refers to the level, and $\mathbf{R}$ are the atomic positions, considered as a parameter here. 

The minimum energy geometry for the ground- and excited-levels will differ. We will denote $\mathbf{R}^{(g)}$ and $\mathbf{R}^{(e)}$ as the \textit{relaxed} positions of the ground and excited states, respectively. Taking the coordinates of the ground state as reference ($\mathbf{R}^{(g)}$), and expanding to the first order:
\begin{eqnarray}
H(\mathbf{R}^{(e)}) &=& \sum_{i=1}^2 E_i(\mathbf{R}^{g})c_i^\dagger c_i 
+ \sum_{i=1}^2\sum_{j=1}^{3N} \nabla_jE_i(\mathbf{R}^{g})\Delta \mathbf{R}_j c_i^\dagger c_i \nonumber\\
&=& H_0 + H_{eph},
\end{eqnarray}
\noindent with $\Delta \mathbf{R} = \mathbf{R}^{(e)}-\mathbf{R}^{(g)}$, and the index $j$ running over all the atomic coordinates.

By using $\Delta \mathbf{R} = a^\dagger+a$ (with $a,a^\dagger$ bosonic operators), the last term takes the form of a typical electron-phonon interaction:
\begin{eqnarray}
    H_{eph} &=& \sum_{i=1}^2\sum_{j=1}^{3N} \mathbf{F}_{i,j}\Delta \mathbf{R}_j c_i^\dagger c_i \nonumber\\
    &=& \sum_{i=1}^2\sum_{j=1}^{3N} g_{i,j} (a_j^{\dagger}  + a_j) c_i^\dagger c_i,
\end{eqnarray}
\noindent where the electron-phonon coupling strength is $g_{i,j}=\mathbf{F}_{i,j}=\nabla_jE_i(\mathbf{R}^{g})$.

It is useful to switch to the basis of the phonons of the system:
\begin{eqnarray}
    \mathbf{F}_{i,k} &=& \sum_{j=1}^{3N}\mathbf{F}_{i,j}\Delta\mathbf{r}_{k,j} \label{Fk}\\
    \Delta\mathbf{R}_k &=& \sum_{j=1}^{3N}\Delta\mathbf{R}_{j}\Delta\mathbf{r}_{k,j} \label{Rk}
\end{eqnarray}
with $\Delta\mathbf{r}_{k,j}$ the $j$ coordinate of $k$-th (normalized) phonon mode. In this basis the electron-phonon Hamiltonian is:
\begin{equation}
    H_{eph} = \sum_{i=1}^{2}\sum_{k=1}^{3N}g_{i,k}\left(b_k^\dagger+b_k\right)c_i^\dagger c_i,
\end{equation}
with $b_k, b_k^\dagger$ bosonic operators acting on the phonon modes. 

Finally, the full Hamiltonian, $H_t$ (adding the phonon energy) reads:
\begin{eqnarray}
    H_{t} &=& \sum_{i=1}^2 E_i(\mathbf{R}^g)c^\dagger_i c_i
    + \sum_{i=1}^{2}\sum_{k=1}^{3N}g_{i,k}\left(b_k^\dagger+b_k\right)c_i^\dagger c_i \nonumber\\
    &+& \sum_{k=1}^{3N} \omega_k b^\dagger_k b_k
\end{eqnarray}

Among the $3N$ phonons involved, only those involved in the atomic rearrangements due to the (de)excitation play a relevant role, \textit{i.e.} a non-negligible $g_{i,k}$. Fig.~\ref{fig1}d shows prominent phonon replicas at $\hbar \omega_k \approx 160-180$ meV. They correspond to localized optical phonons, mostly modes stretching C--C bonds. From a practical point of view, the values of $g_{i,k}$ can be obtained from DFT calculations, just by calculating the forces of  the excited state in the ground state geometry ($\mathbf{F}_{i,j}$). In most DFT codes the excited state can be obtained by fixing the occupations of the levels, the so-called $\Delta$SCF method. Finally, by writing the forces in the basis of the phonons we obtain $g_{i,k}$. An additional approximation is to assume the same phonon spectrum in both, ground and excited states\cite{Alkauskas_2014,Jara2021}. This leads to drop the index $i$.
\begin{figure*}[ht]
\centering
\includegraphics[width=0.331\linewidth]{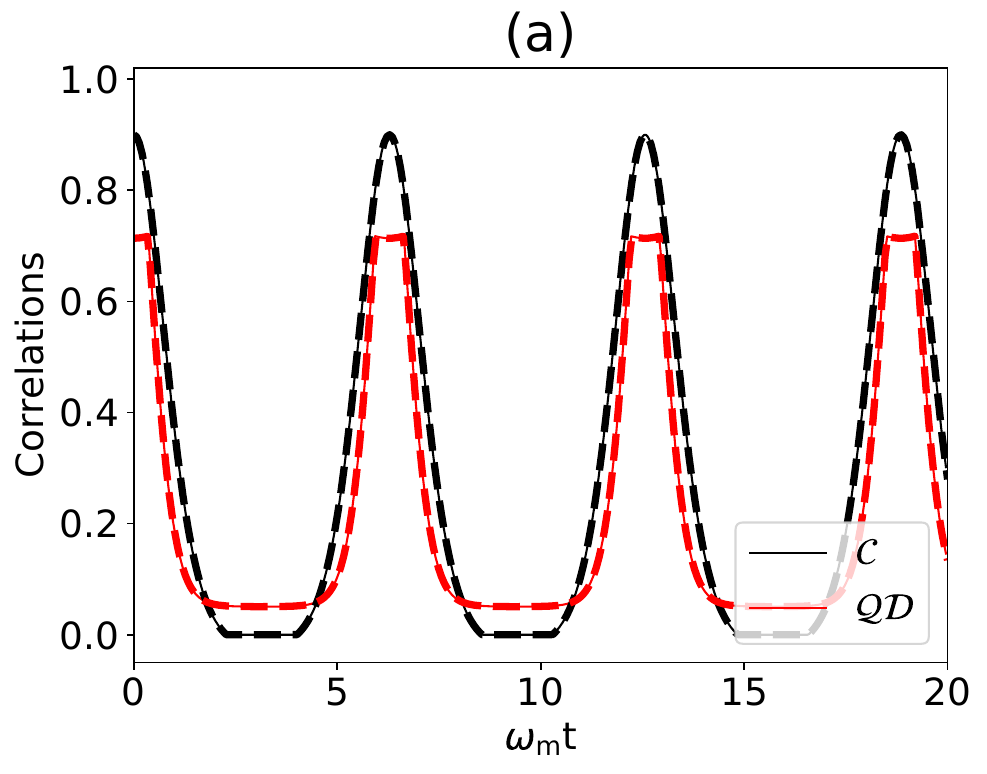}
\includegraphics[width=0.325\linewidth]{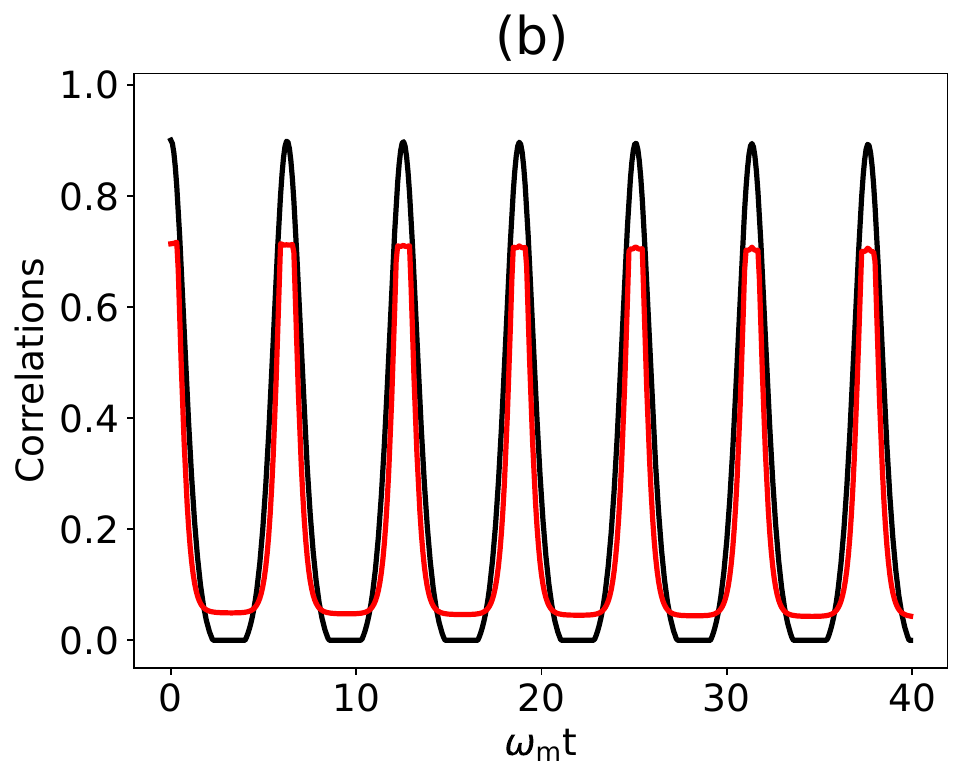}
\includegraphics[width=0.325\linewidth]{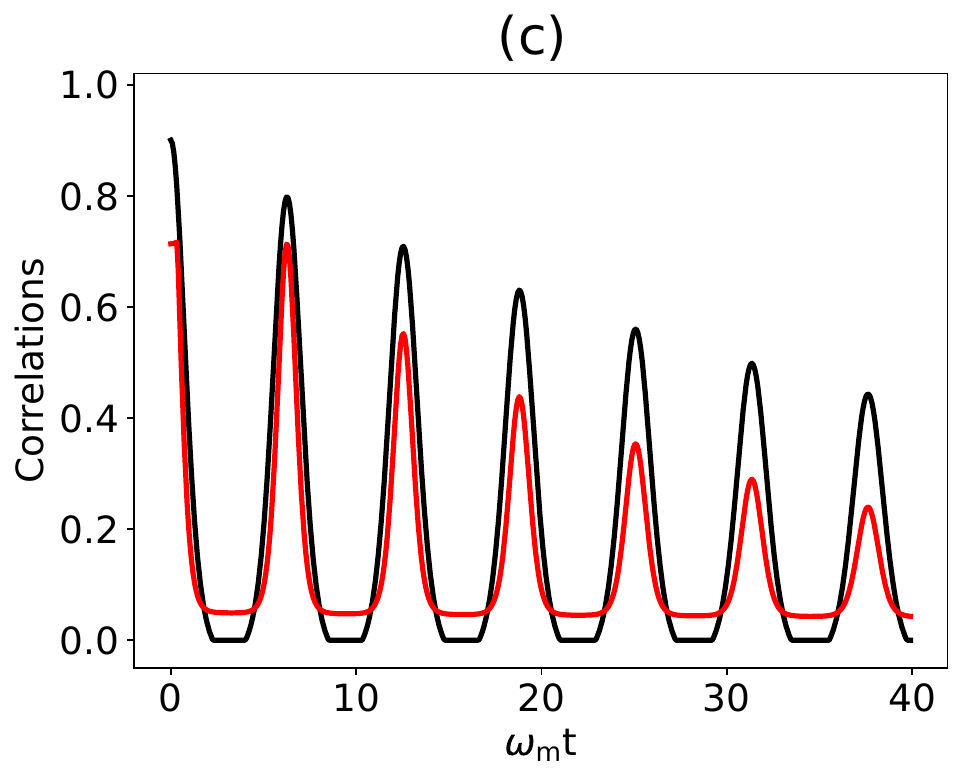}
\caption{Dynamics of Quantum Discord ($\mathcal{QD}$) and Concurrence ($\mathcal{C}$) for SPEs initialized in a BD state (\ref{BDS}). $(a)$ Lossless dynamics considering the numerical simulation of Eq.~(\ref{GME}) (solid line), and analytical solution in Eq.~(\ref{operator}) (dashed line). $(b)$ Dynamics in case of TPL ($\gamma_{b}=0$) with losses for SPEs as $\gamma_{s}=\gamma_{\phi}=10^{-5}$. $(c)$ Dynamics in case of non-topological phonon ($\gamma_{b}=10^{-2}$) with the losses for SPEs as in (b). Other parameters are the same as in Fig.~\ref{fig4} with additional ones: $c_{1}=1$, $c_{2}=-0.9$, $c_{3}=0.9$.}
\label{fig9}
\end{figure*}
Since our interest is only in the phonon modes belonging to the topologically-protected line, we can ignore the other values of $k$. Fig.~\ref{fig:Fk} shows the forces induced by the excitation. Several phonon modes are being excited, including those of the topological line. The value of $g_k=F_k$ for the geometry of Fig.~\ref{fig1}e is $g_k\approx 0.02$ meV\AA. Finally, to get a coupling strength $g_{i,k}$ in units of eV, we will scale it by the lattice parameter of hBN, $a\approx2.5$~\AA. 

The other parameters can be obtained by experimental values or estimated from DFT calculations. The zero-phonon line (\textit{i.e.} the energy of the single photons,  can be obtained from the literature\cite{Tran2016,jero216}, and it ranges from the near-infrared and the near-UV. In $H_t$, $E_i$ refers to vertical transitions, which differ in 0.1-0.2 eV with the measured zero-phonon-line. We used a typical value of $E_i=2.1$ eV. The energy of the (topological) phonons localized at the grain boundary, $\sim 150$ meV. 

\section{Analytical calculation of the density matrix for the Bell-diagonal state}
\label{appQD}
In the $SU(2)$ basis of the Pauli matrices $\hat{\sigma}^{i}$ ($i=\{1,2,3\}\equiv\{x,y,z\}$), defined by an X-type density matrix in Bloch form, the Bell-diagonal (BD) state for the system of two SPEs is
\begin{equation}
    \hat{\rho}_{BD}(0)=\frac{1}{4}\left(\hat{\mathbb{1}}_{2} \otimes \hat{\mathbb{1}}_{2}+\sum_{i=1}^{3}c_{i}\hat{\sigma}^i\otimes\hat{\sigma}^i\right), 
    \label{BDS}
\end{equation}
where $c_{i}$ are real constants satisfying the constraints $ -1\leq c_i \leq 1$, such
that $\hat{\rho}$ is a well defined density operator; $\hat{\mathbb{1}}_2$ is the identity operator.
Since in the initial state the SPEs and phonon are disentangled (no previous interaction), the density matrix of the whole system can be written as a tensor product between the subsystems:
\begin{equation}
   \hat{\rho}(0)=\hat{\rho}_{BD}(0)\otimes\hat{\rho}_{b}(0),
\end{equation}
where $\hat{\rho}_{b}(0)$ is the initial phonon state defined in Eq. \ref{inphon}. 

Now, for the lossless case it is possible to get the analytical dynamics of the SPEs initialized in a BD state. Therefore, considering the Hamiltonian \ref{Hint} without the driving term (\textit{i.e.} $\Omega=0$), in a rotating frame at the SPE frequency one has
\begin{equation}\label{hnew}
    \mathcal{\hat{H}}_{int} = \hat{b}^{\dagger}\hat{b}+g_{0}\left(\hat{\sigma}^z_1+\hat{\sigma}^z_2\right)\left(\hat{b}+\hat{b}^\dagger\right).
\end{equation}
Then the unitary evolution operator reads 
\begin{eqnarray*}
\mathcal{\hat{U}}(t) &=& \exp{\left[-it\mathcal{\hat{H}}_{int}\right]}\\
&=&\exp{\left[g_{0}(\hat{\sigma}^z_1+\hat{\sigma}^z_2)(\eta \hat{b}^{\dagger}-\eta^{*}\hat{b})\right]}\exp{\left[-i\hat{b}^{\dagger}\hat{b}t\right]},    
\end{eqnarray*}
where $\eta=1-e^{-it}$.  In Eq. \ref{hnew} we have normalized the parameters by $\omega_{m}$, i.e. $g_{0}=g_{j}/\omega_{m}$ and $\omega_{0}=\omega_{j}/\omega_{m}$. 

The dynamics of the state of two SPEs is given by the density operator $\hat{\rho}_{s_{1},s_{2}}(t) = Tr_{b}\left\{\mathcal{\hat{U}}(t)\left(\hat{\rho}_{BD}(0)\otimes\hat{\rho}_{b}(0)\right)\mathcal{\hat{U}}^{\dagger}(t)\right\}$.
    
After a straightforward calculation, one obtains the two-qubit state in a $X$-form
\begin{equation}\label{operator}
\hat{\rho}_{s_{1},s_{2}}(t)=\frac{1}{4}
    \begin{pmatrix}
        1+c_{3} & 0 & 0 & \left(c_{1}-c_{2}\right) A\\
        0 & 1-c_{3} & c_{1}+c_{2} & 0\\
        0 & c_{1}+c_{2} & 1-c_{3} & 0\\
        \left(c_{1}-c_{2}\right)A & 0 & 0 & 1+c_{3}
    \end{pmatrix}
\end{equation}
where $A=Tr_{b}\left\{D(4g_{0})e^{-i\hat{b}^{\dagger}\hat{b}t}\hat{\rho}_{b}(0)e^{i\hat{b}^{\dagger}\hat{b}t}\right\}$, 
with $D(4g_{0})=e^{4g_{0}\left(\eta \hat{b}^{\dagger}-\eta^{*}\hat{b}\right)}$.
Since we are studying unitary dynamics here, then the density matrix in Eq.~\ref{operator} retains its X-form during evolution and for example, the Quantum Discord ($\mathcal{QD}$) can be calculated analytically based on the detailed explication in Ref.~\cite{Alber10} (see Eqs. 23-26 therein), where the off-diagonal marginal elements in Eq.~\ref{operator} become time dependent in our case, according to the time function of $A$ defined above. 

\section{Quantum correlations witnessed by Discord}
\label{appQDNew}

The quantum correlations beyond the Entanglement can be witnessed by another quantity known as Quantum Discord ($\mathcal{QD}$) \cite{Henderson2001, Ollivier2002, Luo2008}, and sometimes it is used as an alternative quantum resource in quantum information technologies \cite{Datta2008, Lanyon2008, Dakic2012, Coto2017}. The main feature of $\mathcal{QD}$ is to quantify the quantum correlations for the separated states, \textit{i.e.} those not entangled, and to discriminate which of these states contains quantum (not limited to entanglement) or pure classical correlations.  
For a pedagogical overview of the concepts and definitions related to $\mathcal{QD}$, we recommend a mini-review \cite{Modi2014}, and for more comprehensive material the review \cite{Bera2018}. On the other hand, we point out that the analytical calculation of the $\mathcal{QD}$ is not a straightforward calculation, but it is generally considered a difficult mathematical problem because it involves a minimization (maximization) procedure over many parameters. However, there are few particular cases for which the $\mathcal{QD}$ can be calculated relatively easily, one of these is the case of the Bell-diagonal (BD) states \cite{Luo2008, Alber10,Horodecki4}. 

Let us study the dissipative evolution of SPEs for an initial BD state, which exhibits so called ``\textit{Quantum Discord Freezing}'' (QDF) effects, similar to those studied in \cite{Maziero2009,Mazzola2010,Xu2011,Eremeev14} and other works. In order to analyze such a ``freezing'' effect for $\mathcal{QD}$, we consider the initial BD state with $\{c_i\}=\{1,-0.9,0.9\}$, which has a non-zero Entanglement ($\mathcal{C}$) and non-zero $\mathcal{QD}$. For this quantum protocol, we propose to study three different situations: \textit{i)} Unitary dynamics (no losses at all); \textit{ii)} Dynamics for TPL with losses of SPEs; \textit{iii)} Dynamics for non-topological phonon ($\gamma_{b}=10^{-2}$) with losses of SPEs. The losses of SPEs are defined by $\gamma_{s}=\gamma_{\phi}=10^{-5}$ for a numerical calculation of Eq. \ref{ME2}. In addition to the dynamics of $\mathcal{QD}$, we compute the Concurrence for the same situations. The unitary dynamics of $\mathcal{QD}$ is calculated analytically as explained in Appendix \ref{appQD}, whilst the dissipative dynamics can be computed numerically by solving the general ME, Eq.~\ref{ME2}.

As a result, in Fig. \ref{fig9} the evolution of the initial two-qubit BD state exhibits an interesting effect, since the $\mathcal{QD}$ is frozen periodically in time (plateaux regions). In contrast, the Concurrence does not evidence such effect, only exhibiting the so-called ``\textit{Entanglement Sudden Death}'' effect (the regions where $\mathcal{C}=0$). In Fig.~\ref{fig9}a we compare the numerical and analytical (Eq.~\ref{operator}) results, that fit together perfectly for the unitary dynamics of the full system. When the thermal losses are included into the dynamics, the effect of freezing is destructively affected in time, as observed in Fig. \ref{fig9}(b-c). However, for the case of TPL as compared to the non-TPL case, the freezing is protected for a longer time period, as seen by comparing Figs. \ref{fig9}b and \ref{fig9}c. 

As the main conclusion of this subsection, we point out that the $\mathcal{QD}$ can be generated and the QDF effect observed for a particular type of BD states, in which the qubits (SPEs here) are initialized, and for a certain range of values of the model parameters, such as the relationship between coupling and decoherence rates within the subsystems.


\section{Quantum circuit for the IBM Quantum Simulator}
\label{appIBM}

\begin{figure}[h]
\centering
\includegraphics[width=1\linewidth]{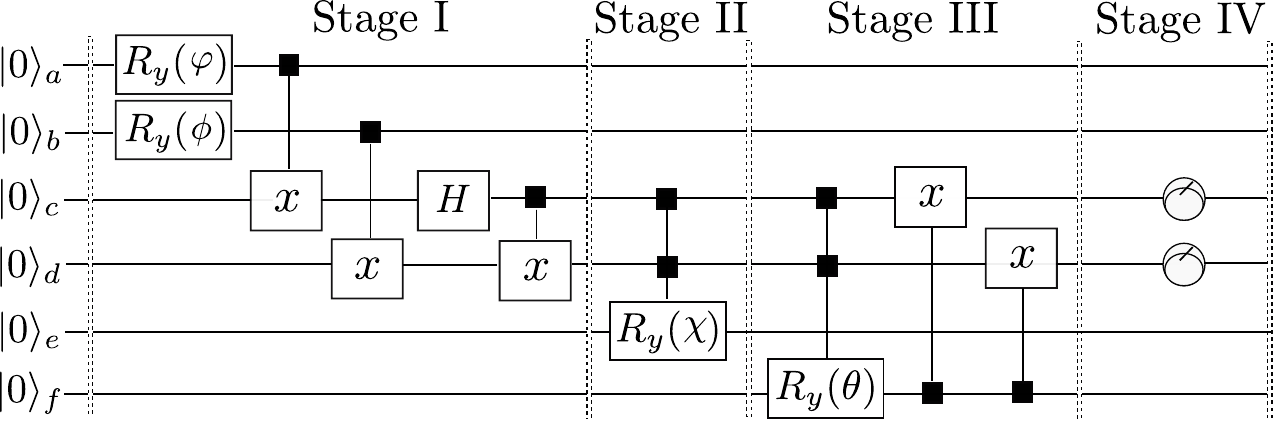}
\caption{Quantum circuit simulating the dissipative evolution of the two SPEs interacting with TPL under the losses mechanisms of the SPEs. This circuit emulates the quantum evolution of the whole system equivalent to the ME \ref{ME2}.}
\label{fig10}
\end{figure}

In order to realize a quantum experiment on IBM Qiskit simulator one needs to emulate evolution of the system and measurement of the interested quantities via a quantum circuit running on the simulator. Therefore, in Fig. \ref{fig10} we show the quantum circuit that simulates the open dynamics of the two dissipative SPEs interacting via TPL and governed by the ME, Eq. \ref{ME2}. 

The quantum circuit involves several operational stages as can be observed in Fig. \ref{fig10}. Following Refs. \cite{Pozzobom2019,e23070797}, it is possible to prepare the SPEs in Bell diagonal states (see \emph{stage I}). Here, the single-qubit rotation about the $Y$ axis, \emph{$R_{y}(\varphi)$}  and \emph{$R_{y}(\phi)$}, are applied on the lines $|0\rangle_{a}$ and $|0\rangle_{b}$, respectively. The single-qubit rotation about the $Y$ axis is generally defined for an angle $\vartheta$ as:
\begin{equation}\label{ry}
    R_{y}(\vartheta)= e^{-\imath\frac{\vartheta}{2}\hat{\sigma}^{y}}=
    \begin{pmatrix}
    \cos{(\vartheta/2)} & -\sin{(\vartheta/2)}\\
    \sin{(\vartheta/2)} & \cos{(\vartheta/2)}\\
\end{pmatrix}.
\end{equation}
\begin{figure*}[ht]
\centering
\includegraphics[width=0.36\linewidth]{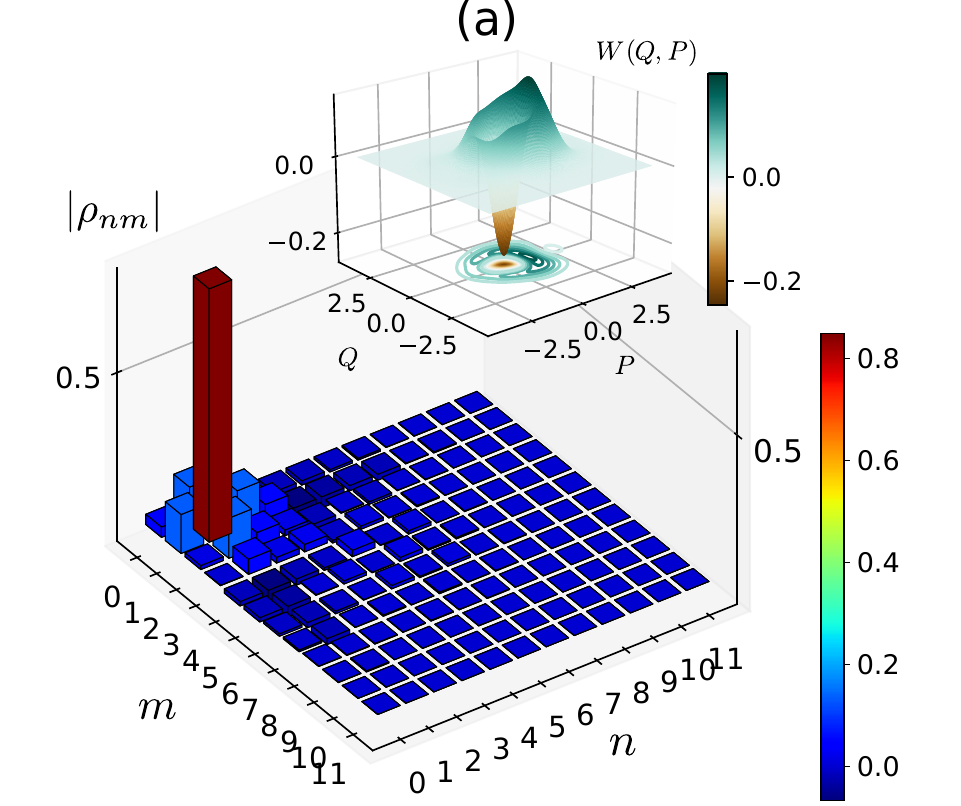}
\includegraphics[width=0.36\linewidth]{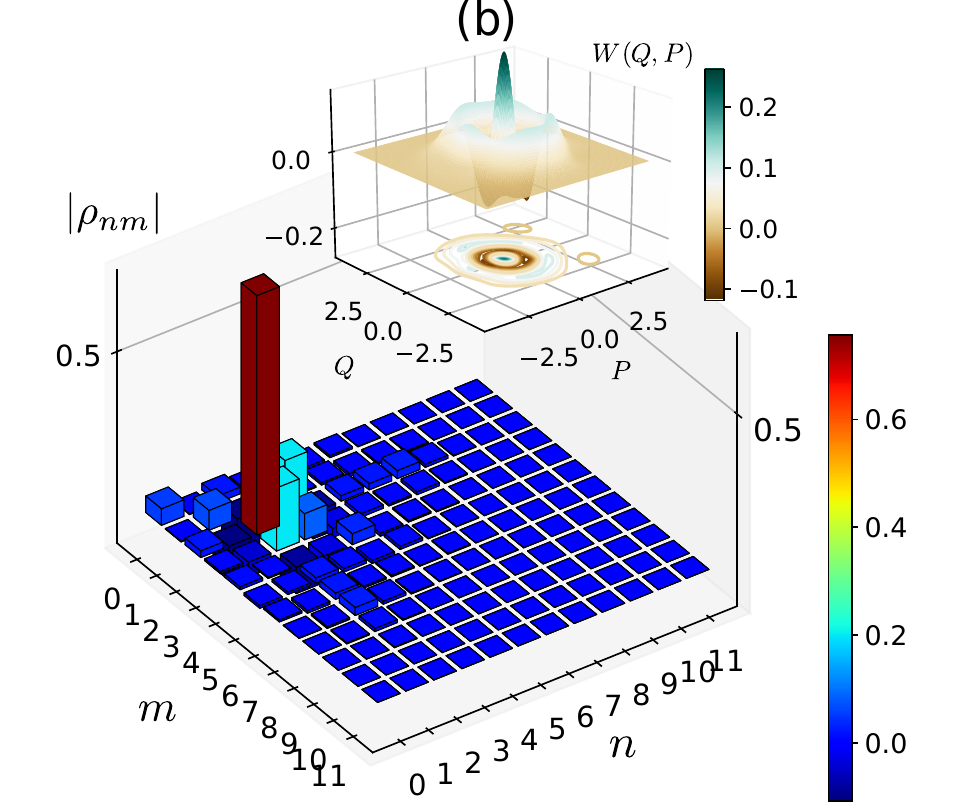}
\vspace{0.5cm}

\includegraphics[width=0.36\linewidth]{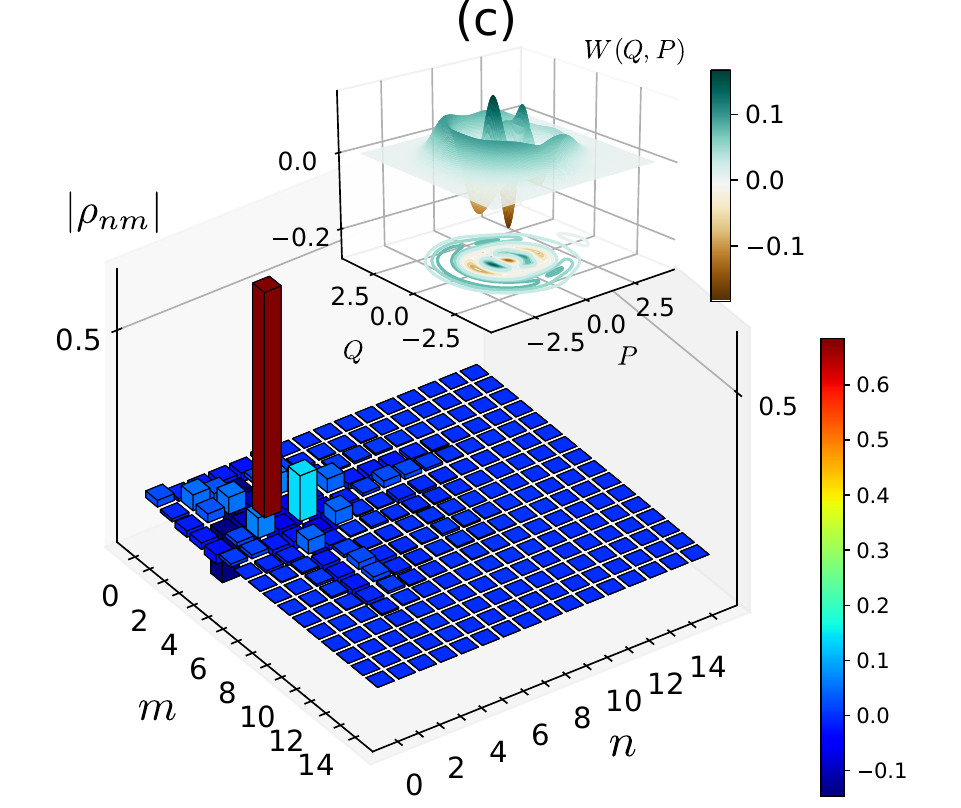}
\includegraphics[width=0.365\linewidth]{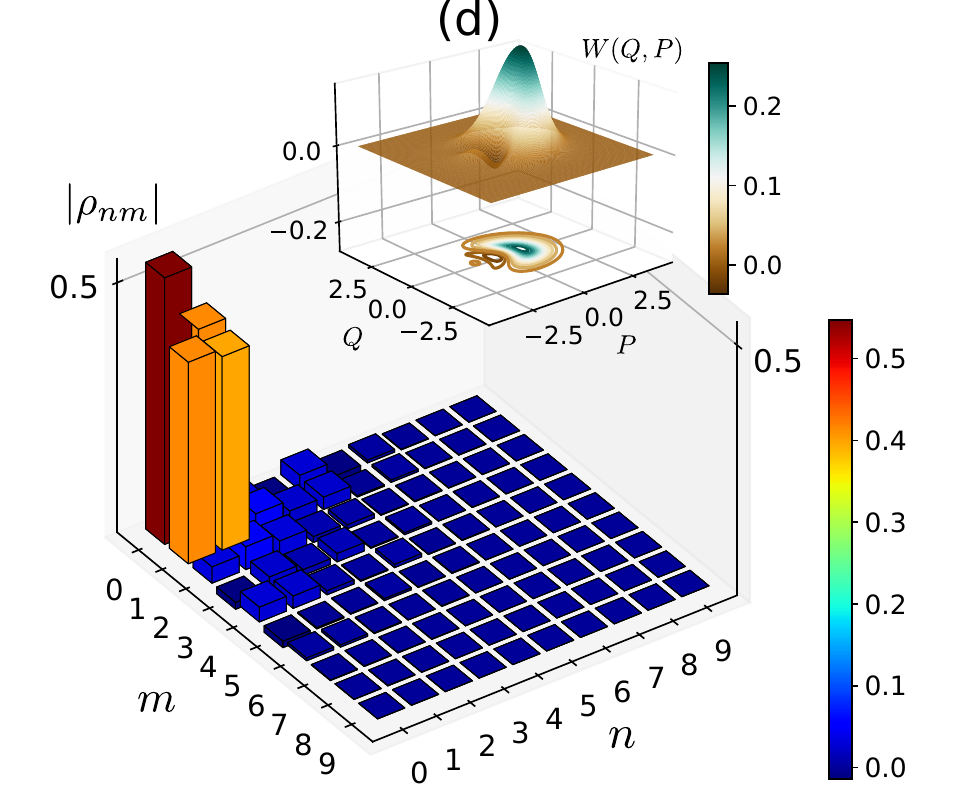}
\caption{Generation of phonon Fock and qubit states for two equivalent SPEs, i.e. $\Delta_{1}=\Delta_{2}=\Delta$, $\Omega_1=\Omega_2=\Omega$,  $\alpha_1=\alpha_2=\alpha$, and $g_1=g_2=0.33\omega_m$. $(a)$ Single-phonon state with a fidelity $\mathcal{F}_{\vert 1 \rangle}=0.92$ obtained for $\Delta=0.33$, $\Omega=0.71$, and $\alpha=0.15$. $(b)$ Phonon Fock state $\vert 2 \rangle$ with a fidelity $\mathcal{F}_{\vert 2 \rangle}=0.88$ for $\Delta=0.83$, $\Omega=0.53$, and $\alpha=0$ ; $(c)$ Phonon Fock state $\vert 3 \rangle$ with a fidelity $\mathcal{F}_{\vert 1 \rangle}=0.82$ for $\Delta=1.73$, $\Omega=1.26$, and $\alpha=0.46$; $(d)$ Phonon-qubit state with a fidelity $\mathcal{F}_{qubit}=0.94$ for $\Delta=0.33$, $\Omega=0.33$, and $\alpha=0.96$. (inset: Negative values of Wigner function witnesses a quantum state). Other parameters are the same as in Fig.~\ref{fig2}}
\label{fig11}
\end{figure*}
The aforementioned phases, $\varphi$ and $\phi$ can be obtained using the expressions \cite{Pozzobom2019,e23070797}:
\begin{align}
\varphi&=2\arccos{\sqrt{p_{00}+p_{01}}},\\
\phi&=2\arccos{\sqrt{p_{00}+p_{10}}}.
\end{align}
Here $p_{00}, p_{01}, p_{10}, p_{11}$ are the parameterization between Pauli and Bell basis $\left\{p_{jk}\right\}_{j,k=(0,1)}\rightarrow\left(c_{1},c_{2},c_{3}\right)$ given by
\begin{align}
    p_{00}&=\frac{1+c_{1}-c_{2}+c_{3}}{4},\\
    p_{01}&=\frac{1-c_{1}+c_{2}+c_{3}}{4},\\
    p_{10}&=\frac{1+c_{1}+c_{2}-c_{3}}{4},\\
    p_{11}&=\frac{1-c_{1}-c_{2}-c_{3}}{4}.
\end{align}

In the following one introduces the operations X-control (\emph{X}) and Hadamard (\emph{H}) defined as \cite{Nielsen}:
\begin{equation}
    X=
    \begin{pmatrix}
    1 & 0 & 0 & 0\\
    0 & 0 & 0 & 1\\
    0 & 0 & 1 & 0\\
    0 & 1 & 0 & 0\\
    \end{pmatrix}, \quad
    H=\frac{1}{\sqrt{2}}
    \begin{pmatrix}
    1 & 1 \\
    1 & -1 \\
\end{pmatrix}.
\end{equation}

The \emph{stage II} of the circuit emulates the unitary evolution of the SPE-phonon system. Here, let us consider the evolution of a quantum state $\rho(0)$ under a trace-preserving quantum operation $\rho(t)$ \cite{Wang_2019},
\begin{equation}
    \rho(t)=\sum_{i,j}\left(E_{i}\otimes E_{j}\right)\rho(0)\left(E_{i}\otimes E_{j}\right),
\end{equation}
where $\left\{E_{k}\right\}$ is the set of Kraus operators associated to a decoherence process of a single qubit \cite{Nielsen}.

In our case, the dynamics of the qubit is given by the following Kraus operators
\begin{equation}
    E_{0}(t)=\sqrt{\frac{1+\Theta(t)}{2}}I, \quad E_{1}(t)=\sqrt{\frac{1-\Theta(t)}{2}}\sigma^{z},
\end{equation}
where $\Theta(t)=\sqrt{A}$ defined in Eq.~\ref{operator}. 

To finish this stage, one applies the operation controlled--RX gate \cite{Nielsen} with the angle $\chi=2\arccos{\left[\Theta(t)\right]}$. 

The \emph{stage III} is responsible for the SPEs evolution under the effects of quantum noise. For this, we consider the Amplitude damping channel (AD) with the angle $\theta$ defined as \cite{García-Pérez2020}:
\begin{equation}
    \theta=2\arccos{\left(\sqrt{p_{AD}(t)}\right)}
\end{equation}
where $p_{AD}(t)$ is given by
\begin{equation}
    p_{AD}(t)=e^{-\gamma_s t}\left[\frac{\gamma_s}{r}\sin{\frac{rt}{2}}+\cos{\frac{rt}{2}}\right]^{2}
\end{equation}
where $r=\sqrt{2\gamma_s-\gamma_s^{2}}$ and $\gamma_s$ correspond to the SPEs damping rate as in Eq. \ref{ME2}, and the dephasing is not included here explicitly since it appears as an intrinsic noise of the quantum simulator.

Finally, in \emph{stage IV} the measurement of the SPEs is performed, so we can extract the information from the density operator for SPEs after all the evolution.

\section{Generation of phonon quantum states with two SPEs.}
\label{sec:AppFock1}

Here we present the preparation protocols for phonon Fock and qubit states for the two SPEs case, alternatively to the one-SPE case already considered in the main text. The results are shown in Fig. \ref{fig11}, where the quantum states are generated only when the SPEs system is driven by an external field, i.e. $\Omega \neq 0$, similarly to Figs. \ref{fig2}-\ref{fig3}. 
The protocol of two SPEs exhibits high fidelities almost similar to the phonon Fock states as compared to the case of one SPE.

For the system of two and three SPEs, we can reproduce similar effects to those observed for the Fock and Qubit phonon states in the case of one SPE, but for a different set of parameters $\{\Omega,\Delta,\alpha\}$, see Fig. \ref{fig11} in Appendix \ref{sec:AppFock1}. In particular, we observe that for many SPEs involved in the dynamics, the fidelities for the similar quantum states decreases compared to the case of one SPE. This result is expected, as the total decoherence in the dynamics increases with the number of SPEs and thus affects the purity of the quantum phonon state generation. 

In conclusion, by the action of the coherent driving, it is possible to transfer periodically the quantum coherence from the SPEs to the phonon mode via the interaction defined by $\hat{\mathcal{H}}_I$. As a result, the corresponding phonon mode is built up in a quantum Fock state $\vert n=1 \rangle$, $\vert n=2 \rangle$, etc. On the other hand, without the external driving, \textit{i.e.} $\Omega=0$, it is impossible to get such quantum states, nevertheless one still observes the periodic formation of a coherent phonon state when $g^{(2)}_{b}(0) \rightarrow 1$, see orange curve in Fig. \ref{fig2}a).

%

\end{document}